# On the molecular origins of the ferroelectric splay nematic phase

Richard J. Mandle[1,2*], Nerea Sebastián[3], Josu Martinez-Perdiguero[4] & Alenka Mertelj[3*]

[1] School of Physics and Astronomy, University of Leeds, Leeds, UK, LS2 9JT
[2] Department of Chemistry, University of York, York, YO10 5DD, UK
[3] Jožef Stefan Institute, SI-1000 Ljubljana, Slovenia
[4] Department of Physics, University of the Basque Country (UPV/EHU), Apdo.644-48080 Bilbao, Spain

*Corresponding authors: r.mandle@leeds.ac.uk, alenka.mertelj@ijs.si

**Abstract**

Nematic liquid crystals have been known for more than a century, but it was not until the 60s-70s that, with the development of room temperature nematics, they became widely used in applications. Polar nematic phases have been long-time predicted, but have only been experimentally realized recently. Synthesis of materials with nematic polar ordering at room temperature is certainly challenging and requires a deep understanding of its formation mechanisms, presently lacking. Here, we compare two materials of similar chemical structure and demonstrate that just a subtle change in the molecular structure enables denser packing of the molecules when they exhibit polar order, which shows that reduction of excluded volume is in the origin of the polar nematic phase. Additionally, we propose that molecular dynamics simulations are potent tools for molecular design in order to predict, identify and design materials showing the polar nematic phase and its precursor nematic phases.

# Introduction

Ferroelectric, ferromagnetic and ferroelastic materials hold a privileged position in material sciences and condensed matter physics research. Their remarkable properties (e.g. shape memory effect, pyroelectric effect, large mechanical, magnetic and electric susceptibilities) demonstrate their potential for use in a wide range of novel technologies. Although there is no fundamental reason preventing ferromagnetic or ferroelectric order in dipolar liquids, i.e., liquids made of constituents with dipole moment, it was not until very recently, that three different liquid systems with ferroic order have been realized experimentally[1–3]. Because dipolar interactions are strongly anisotropic, suitable (anisotropic) positional correlations between nearest neighbours are crucial for the appearance of ferroic order[4]. In a liquid, in which the constituents move and rotate, the positional correlations are short-ranged – they span to a few nearest neighbours, and they strongly depend on the shape of the constituents. It seems that the deciding factor for the appearance of long-range ferroic order is such anisometric shape of the constituents which also leads to the formation of anisotropic liquids, better known as liquid crystals. In the case of a simple uniaxial nematic phase, the anisotropic constituents are on average oriented in the same direction, denoted by a unit vector- the director. However, for a long time, the realization of ferroic nematic phases remained elusive. Not long ago, it has been shown that, as theoretically predicted[5], disk-like shape can lead to uniform ferromagnetic nematic order in colloids of disk-like magnetic particles dispersed both in isotropic liquids[2] and in nematic liquid crystals[1,6], and to helical magnetic order when dispersed in chiral nematic liquid crystals[7,8]. Still, polar nematic phases on bulk molecular materials seemed to be prevented due to entropic and dipolar interactions between constituting molecules.

Very recently, it has been shown that some materials made of elongated molecules, with large longitudinal electric dipole moment and a side group which gives molecules a slight wedge shape, exhibit two nematic phases [9,10]. The high-temperature phase is an apolar uniaxial nematic (*N*), the kind widely exploited in current LCD technology, while the low-temperature phase exhibits ferroelectric ordering on the macroscopic scale of several microns [3]. Because the shape of the molecules lacks head-tail symmetry, polar ordering of the molecules causes orientational elastic instability. The transition between the phases is a ferroelectric-ferroelastic transition[3], in which a divergent susceptibility, typical of a ferroelectric transition, is accompanied by the softening of the splay elastic constant which causes the lower temperature ferroelectric phase to be non-uniform[3,11,12], i.e. ferroelectric splay nematic phase *($N_S$)*. As it is impossible to fill the space with homogenous Japanese fan-like splay deformation, the exact structure of the director field in the *$N_S$* will depend on the confinement conditions (size, shape and boundary condition of a container), the orientational elastic constants, the electric polarization and ion concentration. The structure can exhibit 1D[13,14] or 2D[15] modulation, regular or irregular defect lattice[16] or the splay can be combined with other types of deformation, e.g. the twist deformation[17]. Experimentally, a modulated structure with microns size periodicity has been observed[3,11]. It is anticipated that when the material is confined to a layer with a thickness comparable to the periodicity, more metastable structures can occur[12]. The high saturated electric polarization values measured in the typical material *RM734*[18] (Fig.1a) show that materials exhibiting the ferroelectric splay nematic phase are promising for a variety of applications such as low power fast electro-optic switching devices as well as new LC-based photonic technologies, e.g. switchable optical frequency converters. Up to now examples of materials exhibiting this ferroelectric nematic phase are scarce, and limited to some *RM734* analogues[9]. Also, a liquid crystal compound with a 1,3-dioxane unit was reported to exhibit similar textures and polarity in a low temperature nematic phase[19]. Unfortunately, so far the ferroelectric splay nematic phase has only been observed at high temperatures creating a pressing need for the development of new materials which would exhibit the ferroelectric nematic phase at much lower temperatures, suitable for applications. Understanding the microscopic mechanism which leads to the formation of the ferroelectric phase is therefore imperative; such knowledge, combined with the development of predictive algorithms using computational models, would help to tailor the design of new materials that exhibit this phase.

It has been shown that when the nitro group of *RM734* is replaced by a nitrile group (*RM734-CN*, see Fig. 1b), the resulting material does not exhibit the ferroelectric splay nematic phase[9]. Notably, binary mixtures of *RM734* with *RM734-CN* do not show the $N_S$ phase either; even with as little as 10 wt% of *RM734-CN*, the $N_S$ phase is suppressed entirely. In this paper, we aim to gain insight into the driving mechanism for the formation of the $N_S$ phase, by carrying out a comprehensive comparative study of both materials, *RM734* and *RM734-CN*. Experimentally, we focused on those properties that seem to be critical in the $N$-$N_S$ transition, by performing dielectric spectroscopy, dynamic light scattering and WAXS/SAXS experiments. The differences between both materials is then analysed in-depth *via* molecular dynamics simulations.

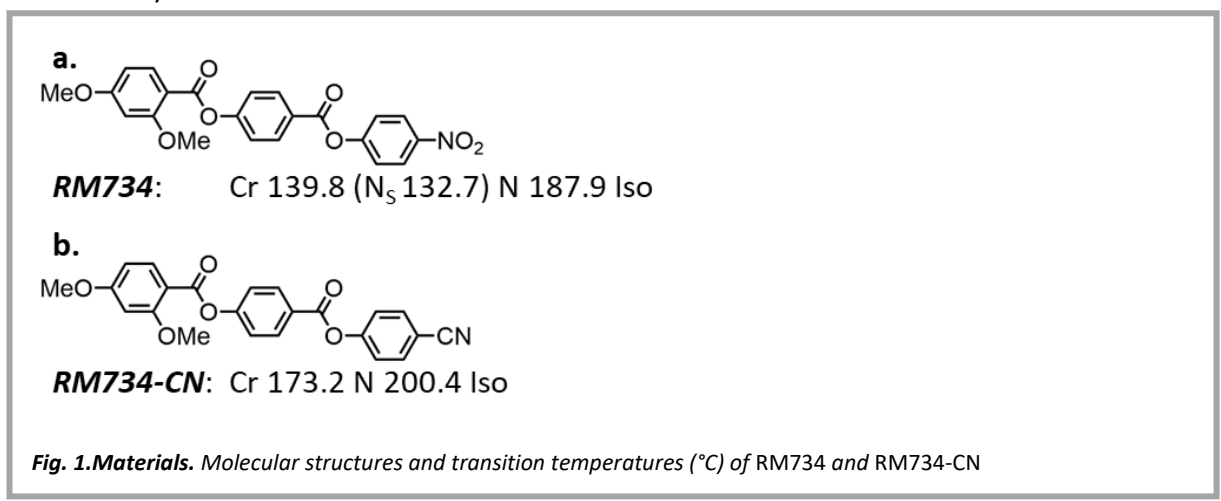

***Fig. 1.Materials.*** *Molecular structures and transition temperatures (°C) of* RM734 *and* RM734-CN

## Results

**Distinct dielectric relaxation spectra**

The dielectric spectrum of the *N* phase of *RM734* is distinctive, being characterized by the gradual emergence of a relaxation mode, whose strength ($\Delta\varepsilon$) strongly increases and its relaxation frequency rapidly decreases when approaching the *N*-*N$_S$* phase transition. Such behaviour is characteristic of increasingly cooperative dipole motions, evidencing the growth of polar correlations in the *N* phase. We have performed the same studies for the analogue *RM734-CN* material in the isotropic and nematic phase (see Materials and Methods).

Similarly to *RM734*, *RM734-CN* spontaneously aligned homeotropically on the untreated gold electrode surfaces, and thus, we measured the parallel component of the permittivity. To ensure no director reorientation during the measurement, the oscillator level was set to 30 mV$_{rms}$ and measurements where performed on cooling. The temperature and frequency dependence of the dielectric spectra is given in Figure SI.1 for the nematic phase. At each temperature the characteristic frequency and amplitude of each relaxation process are obtained by fitting $\varepsilon(\omega)$ to the Havriliak-Negami equation[20]:

$$\varepsilon(\omega) = \sum_k \frac{\Delta\varepsilon_k}{\left[1 + \left(i\omega\tau_{HN}\right)^{\alpha_k}\right]^{\beta_k}} + \varepsilon_\infty - i\frac{\sigma_0}{\omega\varepsilon_0} \qquad (1)$$

Examples of the fits can be found in Fig. 2a and Fig. SI.2. The fit results for the characteristic frequencies and amplitudes of the different relaxation modes obtained for *RM734-CN* are shown in Fig. 2 and compared to those already reported for *RM734*[3]. Isotropic phase is characterized by a broad single relaxation process ($m_{iso}$) with frequency around 40 MHz and amplitude $\Delta\varepsilon_{iso} \sim 12$. On cooling, immediately after the *I-N* transition, dielectric spectra is characterized by a relaxation mode at lower frequencies and with growing amplitude. This mode, although close to a Debye at high temperatures, slowly broadens on decreasing the temperature. This same behaviour was observed for *RM734*. As in the latter case, far below the transition, it becomes evident that the single relaxation mode detected in the high temperature range is indeed composed of two relaxation modes, $m_{\parallel,1}$ and $m_{\parallel,2}$. In line with the observations for *RM734*, the higher frequency one, $m_{\parallel,2}$, can be attributed to molecular rotations around the short molecular axis, while the lower frequency one, $m_{\parallel,1}$, would correspond to the collective reorientation of the dipole moments. However, in the case of *RM734-CN*, dipole correlations are remarkably weaker than in the case of *RM734*. In contrast to the clear softening of the mode for

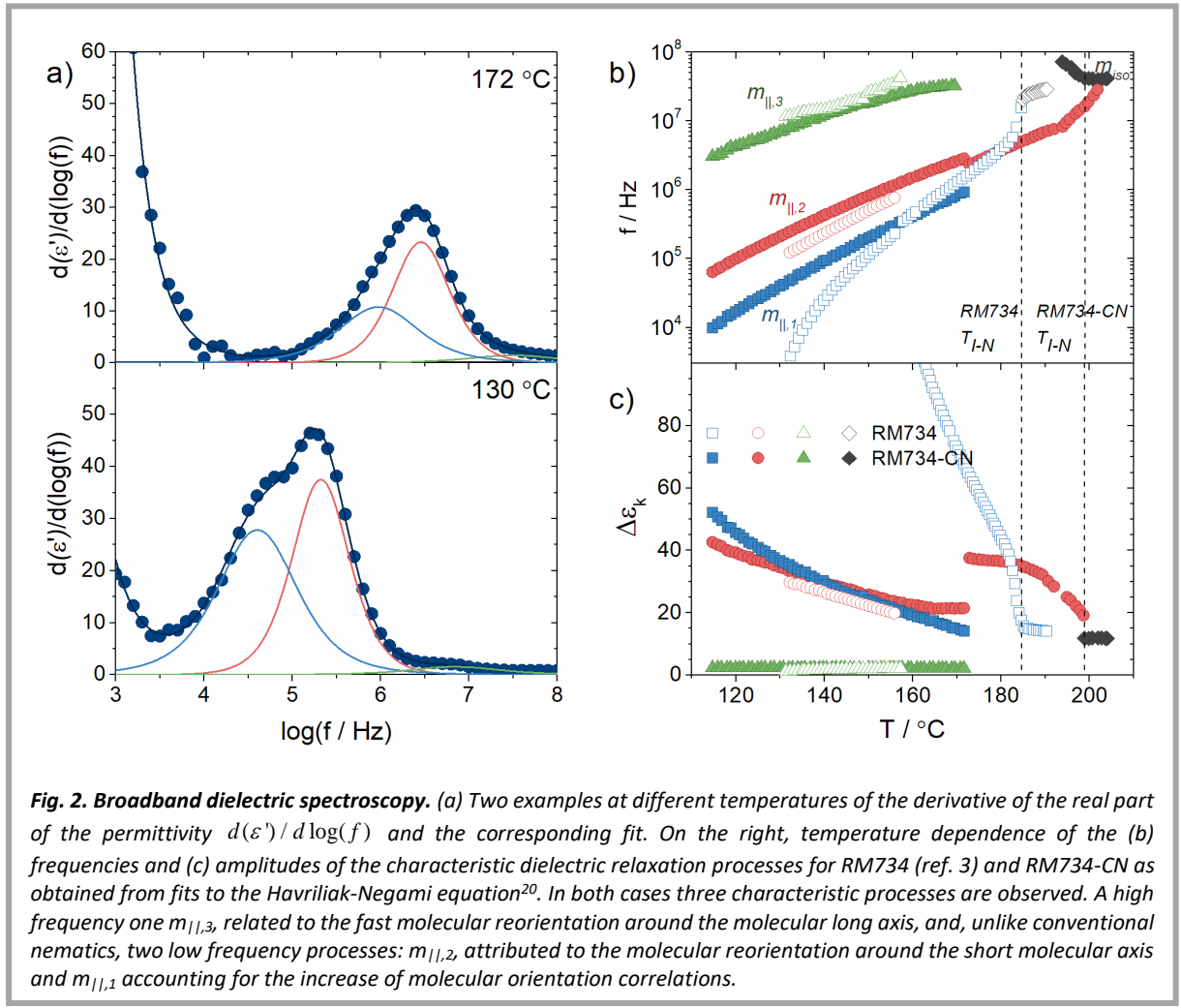

*Fig. 2. Broadband dielectric spectroscopy. (a) Two examples at different temperatures of the derivative of the real part of the permittivity $d(\varepsilon')/d\log(f)$ and the corresponding fit. On the right, temperature dependence of the (b) frequencies and (c) amplitudes of the characteristic dielectric relaxation processes for RM734 (ref. 3) and RM734-CN as obtained from fits to the Havriliak-Negami equation[20]. In both cases three characteristic processes are observed. A high frequency one $m_{||,3}$, related to the fast molecular reorientation around the molecular long axis, and, unlike conventional nematics, two low frequency processes: $m_{||,2}$, attributed to the molecular reorientation around the short molecular axis and $m_{||,1}$ accounting for the increase of molecular orientation correlations.*

*RM734*, the characteristic frequency of $m_{||,1}$ for *RM734-CN* shows Arrhenius-like behaviour in the full *N* temperature range and its strength, although increasing, is far from the diverging trend of *RM734*. In the latter $\Delta\varepsilon_{||,1}$ is much larger than $\Delta\varepsilon_{||,2}$, and the spectrum is dominated by the collective mode. On the other hand, $\Delta\varepsilon_{||,1}$ and $\Delta\varepsilon_{||,2}$ in *RM734-CN* are comparable, indicating that collective reorientations, although present, are weak. It is also interesting to note that the amplitudes of the molecular mode $m_{||,2}$ are comparable in both materials, in agreement with the similar molecular dipole moments of both analogues (see Electronic Structure Calculations section).

In the high range of the measured frequencies (1-10 MHz) a third mode is detected, $m_{||,3}$. At the *I-N* transition, its frequency is larger than that of $m_{iso}$ and rapidly climbs out of the measured frequency range. On further cooling, its characteristic frequency decreases and the mode is again detected. This mode, by frequency and amplitude, can be associated with the rotation around the molecular long axis, as described by the Nordio-Rigatti-Segre theory[21].

**Viscoelastic properties in the N phase**

One more distinctive characteristic of the $N$-$N_S$ transition in *RM734* is the remarkable pretransitional behaviour, characterized by the softening of the splay elastic constant[13]. The ground state of the classical apolar nematic phase is uniform, i.e., the orientation of **n** does not change with position. We measured the orientational elastic constants in *RM734-CN* associated with the increase of the energy due to the splay deformation of **n** using a combination of dielectric Frederiks transition and dynamic light scattering intensities (Materials and Methods[13]) and compared them to those of *RM734*[13]. Comparison of the temperature dependence of the splay elastic constant is shown in Fig. 3a; at high temperatures, $K_1$ of both materials are comparable and unusually low for a nematic liquid crystal. However, it immediately becomes evident that while $K_1$ of *RM734-CN* remains practically constant over the measured *N* range, in the case of *RM734*, the splay elastic constant strongly decreases as the $N$-$N_S$ transition is approached on cooling.

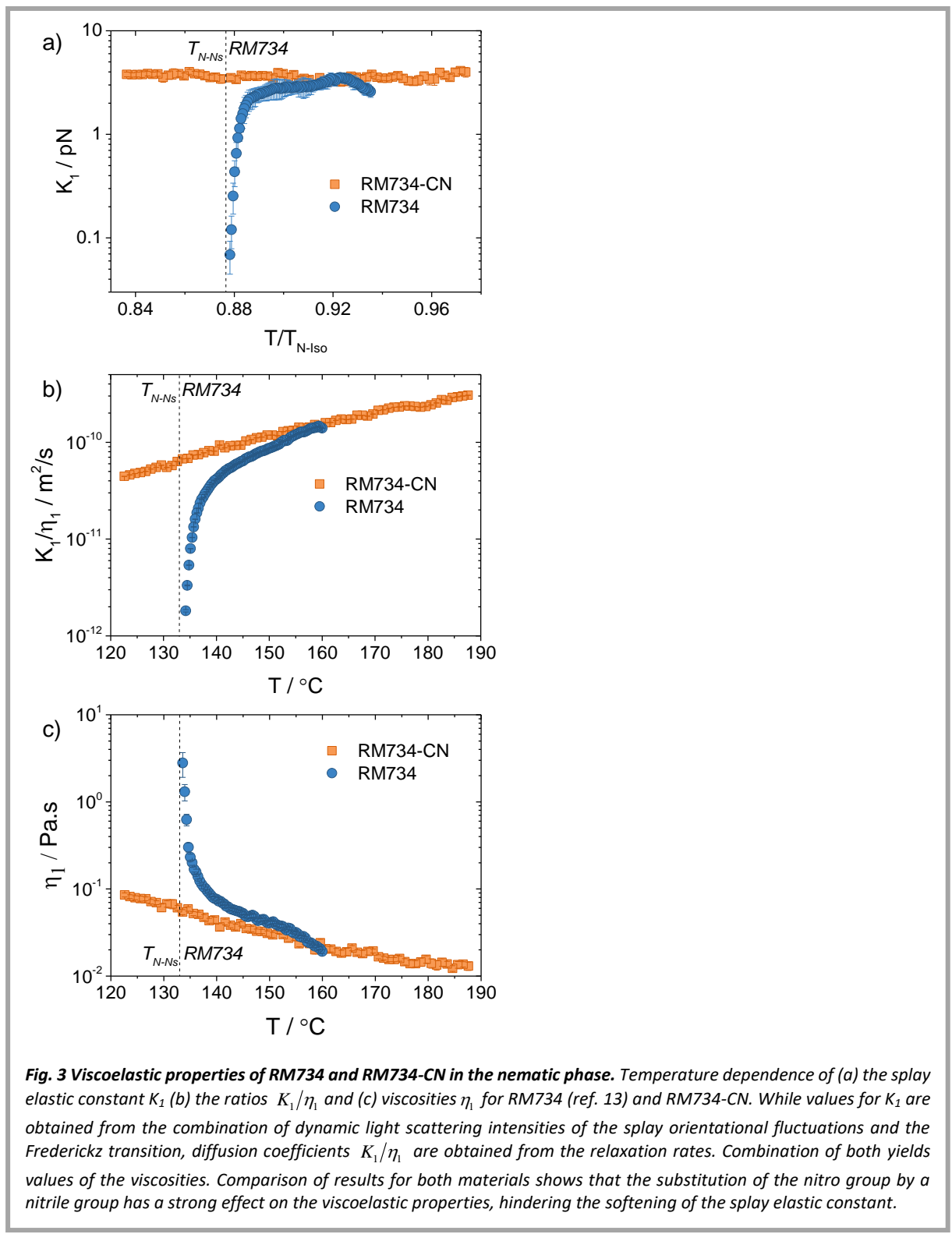

*Fig. 3 Viscoelastic properties of RM734 and RM734-CN in the nematic phase.* Temperature dependence of (a) the splay elastic constant $K_1$ (b) the ratios $K_1/\eta_1$ and (c) viscosities $\eta_1$ for RM734 (ref. 13) and RM734-CN. While values for $K_1$ are obtained from the combination of dynamic light scattering intensities of the splay orientational fluctuations and the Frederickz transition, diffusion coefficients $K_1/\eta_1$ are obtained from the relaxation rates. Combination of both yields values of the viscosities. Comparison of results for both materials shows that the substitution of the nitro group by a nitrile group has a strong effect on the viscoelastic properties, hindering the softening of the splay elastic constant.

Similar contrasting behaviour is observed for the ratio $K_1/\eta_1$ obtained from the relaxation rates of the splay fluctuation modes. As shown in Fig. 3.b, for *RM734-CN* the splay mode becomes slightly slower on lowering the temperature, while a strong softening is detected for *RM734* on approaching the $N$-$N_S$ transition. From the splay elastic constant and the diffusion coefficients $K_1/\eta_1$ we obtained the temperature dependence of the splay viscosity (Fig. 3.c), showing a steep increase before the $N$-$N_S$ transition for the *RM734* as opposed to the classical Arrhenius-like tendency in the case of *RM734-CN*.

**Electronic Structure Calculations**

Thusfar we have shown the dielectric and viscoelastic properties of *RM734* and *RM734-CN* to be distinct; we therefore turned to a combination of electronic structure calculations and molecular dynamic simulations in an attempt to understand the observed differences and how they originate in molecular structure.

An initial investigation into the conformational preference of *RM734* and *RM734-CN* was made with DFT calculations. Starting from geometry optimised at the M06HF-D3/aug-cc-pVTZ level, we performed fully relaxed scans about each of the dihedrals at the same level of theory; *i.e.* the geometry was optimised for each conformation sampled. Selected dihedrals are shown *via* the highlighting of their central bond in Fig. 4. Not unexpectedly, we find that the choice of nitro or cyano terminal unit does not affect the molecular electronic structure in such a way that the conformational preference is affected.

The calculated molecular dipole moments for the global energy minimum geometries of *RM734* and *RM734-CN,* at the M06HF-D3/aug-cc-pVTZ level of DFT, are 11.4 and 11.2 Debye, respectively. For a molecule oriented with its mass inertia axis (taken to be the eigenvector associated with the smallest eigenvalue of the inertia tensor) along x, the dipole vector components are {10.9931,2.9740,0.4941} and {10.7360,2.9549,0.4850}, resulting in an angle between the direction of the dipoles and the molecular axis of 18.3° and 20° for *RM734* and *RM734-CN*, respectively (Fig. SI.5). Additionally, the calculated eigenvalues of polarizabilities at 800nm (given in Table SI.1) at the same level of theory, show that while both molecules have comparable isotropic polarizabilities, the anisotropic polarizability is larger for *RM734-CN* in correspondence with the larger value of the birefringence measured for the nitrile analogue (Fig. SI.3).

We next sought to explore if the distribution of energetically more favourable intermolecular pairing configurations could be among the possible key factors influencing the appearance of the $N_S$ phase. We calculated the counterpoise corrected complexation energy of homogenous pairs of *RM734* and *RM734-CN* at the M06HF-D3/aug-cc-pVTZ level of DFT. The orientation of a geometry optimisation was defined by defining the angle between two molecules as $\omega = \angle\left(\overrightarrow{C_{13}C_{20}}\overrightarrow{C_{67}C_{72}}\right)$, where $C_{12}/C_{67}$ and $C_{20}/C_{72}$ are the distal aromatic carbon atoms within each molecule. During optimisation, no geometry constraints were applied beyond the initial starting configuration, and as the molecules deviate from the starting parallel/antiparallel orientation we present the initial and final orientations in Table 1 and in Fig. 4. In the 'parallel' form of *RM734-CN*, the molecular long axes are notably more rotated with regards to one another ($\omega_{final}$ = 27 °) than for *RM734* ($\omega_{final}$ = 8 °). Additionally, in both cases the long axes are staggered, i.e. their end-to-end positions are shifted. Conversely, relative molecular orientation in the antiparallel forms of the two materials are comparable, and end-to-end positions are aligned. Calculated complexation energies indicate that for both *RM734* and *RM734-CN* the antiparallel form is energetically favoured over the parallel form, although both cases are lower in energy than two individual molecules. The present DFT calculations on pairs of molecules of *RM734* and *RM734-CN* point to both parallel and antiparallel pairing being energetically favourable, with some differences between the geometries of the parallel pairs, but cannot explain the observed differences in dielectric relaxation or viscoelastic properties.

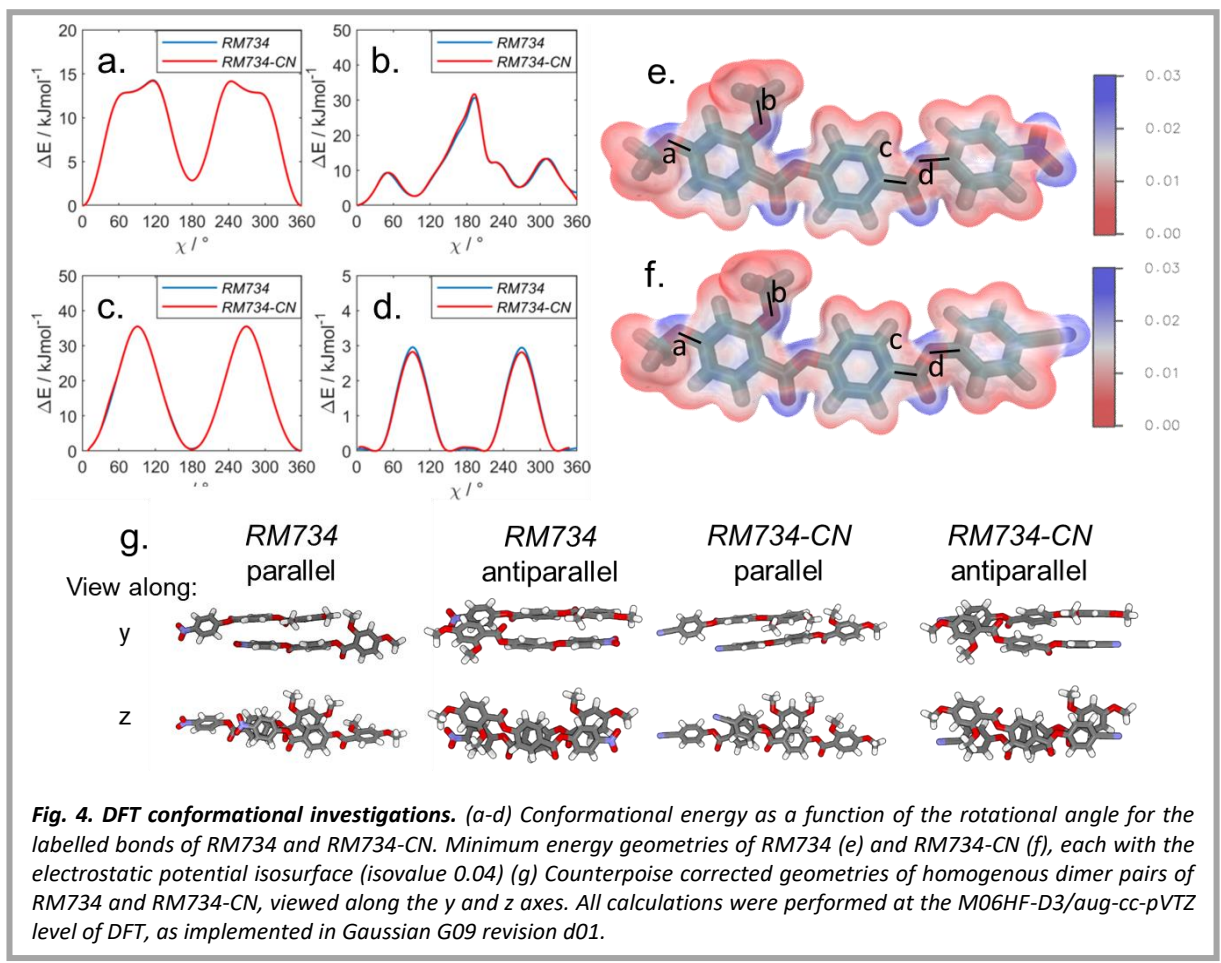

***Fig. 4. DFT conformational investigations.*** *(a-d) Conformational energy as a function of the rotational angle for the labelled bonds of RM734 and RM734-CN. Minimum energy geometries of RM734 (e) and RM734-CN (f), each with the electrostatic potential isosurface (isovalue 0.04) (g) Counterpoise corrected geometries of homogenous dimer pairs of RM734 and RM734-CN, viewed along the y and z axes. All calculations were performed at the M06HF-D3/aug-cc-pVTZ level of DFT, as implemented in Gaussian G09 revision d01.*

| Material | $\omega_{initial}$, ° | $\omega_{final}$, ° | $\Delta E_{int}$, kcal mol$^{-1}$ |
|---|---|---|---|
| *RM734* | 0 | 8 | -19.96 |
|  | 180 | 165 | -34.32 |
| *RM734-CN* | 0 | 27 | -18.12 |
|  | 180 | 166 | -31.39 |

**Table 1:** *Counterpoise corrected complexation angles and energies ($E_{int}$, Kcal mol$^{-1}$) for homogenous parallel and antiparallel dimers of RM734 and RM734-CN.*

**X-ray Scattering, Molecular Dynamics Simulations and Simulated X-ray Scattering Data**

The X-ray scattering behaviour of *RM734* (and that of other structurally related splay nematic materials) is distinct from classical nematic liquid crystals for two principal reasons: the scattering intensity is extremely weak in the liquid and liquid-crystalline states; multiple additional diffuse small angle reflections (002, 003) are present in the *N* and *N$_S$* phases[9]. Neither of these behaviours is observed for the analogous nitrile terminated material, *RM734-CN*, or indeed conventional nematic liquid crystals.

We conjecture that the occurrence of the *N$_S$* phase and the observation of multiple small angle X-ray scattering peaks may be related; specifically, that the additional diffuse small angle reflections observed in the *N* phase of *RM734* (and other materials) encode information about polar order, and so can be used to identify those *N* phases which potentially can lead to the formation of the *N$_S$* phase. To demonstrate this, we sought to calculate/simulate X-ray scattering patterns for polar and apolar nematic states of both *RM734* and *RM734-CN*.

Computation of X-ray scattering intensities for isolated pairs of molecules (e.g. those from DFT calculations, Fig. 4) lacks a description of the nematic structure factor and so cannot explain the observed differences between *RM734* and *RM734-CN*, i.e. the presence of (002) and (003) peaks in the former but not the latter. We, therefore, turned to molecular dynamics (MD) simulations as a means to generate polar (parallel) and apolar (antiparallel) nematic phases comprised of *RM734* and *RM734-CN*, with these then used for subsequent calculation of X-ray scattering intensities.

We performed MD simulations of *both materials* in both polar and apolar starting configurations, at a range of temperatures (see Materials and Methods). Each simulation was maintained at a constant pressure of 1 Bar with anisotropic pressure coupling, enabling the box dimensions to vary. The production MD run of each simulation was 250 ns, as shown in the snapshots in Fig. 5a. We do not observe a splay modulated structure in our simulations; however this is not unexpected given the modulation period of the $N_S$ phase has been measured to be on the order of several microns[3,11], whereas the final dimensions of each simulation box are approximately 10 x 6 x 6 nm$^3$.

We confirmed the nematic or isotropic nature of each simulation by calculating the second-rank orientational order parameter (*P2*) and biaxial order parameter (*B*) for all trajectory frames recorded during the production MD run, according to equation (2); N is the total number of molecules (680 in all simulations), m is the molecule number within a given simulation.

$$Q_{\alpha\beta} = \frac{1}{N} \sum_{m=1}^{N} \frac{3a_{m\alpha}a_{m\beta} - \delta_{\alpha\beta}}{2} \qquad (2)$$

Where *α* and *β* indicate two Cartesian directions, $a_m$ is the long axis of the mth molecule in the simulation, determined from its inertia tensor and $\delta_{\alpha\beta}$ is the Kronecker delta function. $Q_{\alpha\beta}$ is diagonalized to give three eigenvalues. The order parameter *P2* corresponds to the largest eigenvalue of $Q_{\alpha\beta}$, and the biaxial order parameter *B* corresponds to the difference between the two smallest eigenvalues. The polar order parameter, *P1*, was calculated according to:

$P1 = \langle cos\theta \rangle$ \qquad (3)

where $\theta$ is the angle between the molecular axis and the local director. Calculated values of *P2* are given in Fig. 5b. as a function of temperature together with the obtained density values, the latter in particular offering a critical observation. We find that, at a given temperature, the density of *RM734-CN* is invariant in the polar and apolar nematic simulations (Fig. 5c). Conversely, we find that, at a given temperature, the density of *RM734* is somewhat (~ 0.5 %) larger in the polar nematic than in the apolar. We are encouraged that the simulated density of *RM734* is around 15% less than measured in the solid state for this material by X-ray diffraction (1.473 g cm$^3$; see CCDC deposition number 1851381)[13].

For both *RM734* and *RM734-CN* we find *P2* to be nematic like (i.e. ≥ 0.3) at and below 450 K, and isotropic (i.e. ≤ 0.3) at and above 500 K. This compares favourably with experimentally determined clearing points (*RM734* = 461 K, *RM734-CN* = 473.6 K). The obtained values of *P2* were compared with experimental values obtained *via* WAXS for both *RM734* and *RM734-CN* (Table 2) at a temperature of 400 K. It is immediately apparent that both polar and apolar simulations yield an orientational order parameter which is larger than that obtained experimentally; simulations give mean values of ~0.75 in all four cases which compare with experimentally obtained values of 0.68 and 0.62 for *RM734* and *RM734-CN* respectively[22]. This overestimation of orientational order parameters is consistent with other fully atomistic molecular dynamics simulations[23–25]. In all four simulations the biaxial order parameter, *B*, was negligible (<0.05). As expected, the polar order parameter (*P1*) is large (~ 0.9) for simulations in the polar configuration and negligible (< 0.05) for apolar simulations. As mentioned above, molecular axis and dipole moment of each molecule are not parallel, and thus, we can also calculate the order parameter *P1* of the polarization vector, i.e. *P1*(dipole), as the total dipole moment of the simulation box divided by the sum of the dipoles of the individual molecules (Table 2). Difference between *P1* and *P1(dipoles)* arises from the fact that the molecular axis and the molecular dipole are not parallel. Additionally, we can define the polarization vector (P) as the total dipole of the simulation box divided by its volume and calculate the angle between the director **n** and the direction of the polarization described by the unit vector **np**.

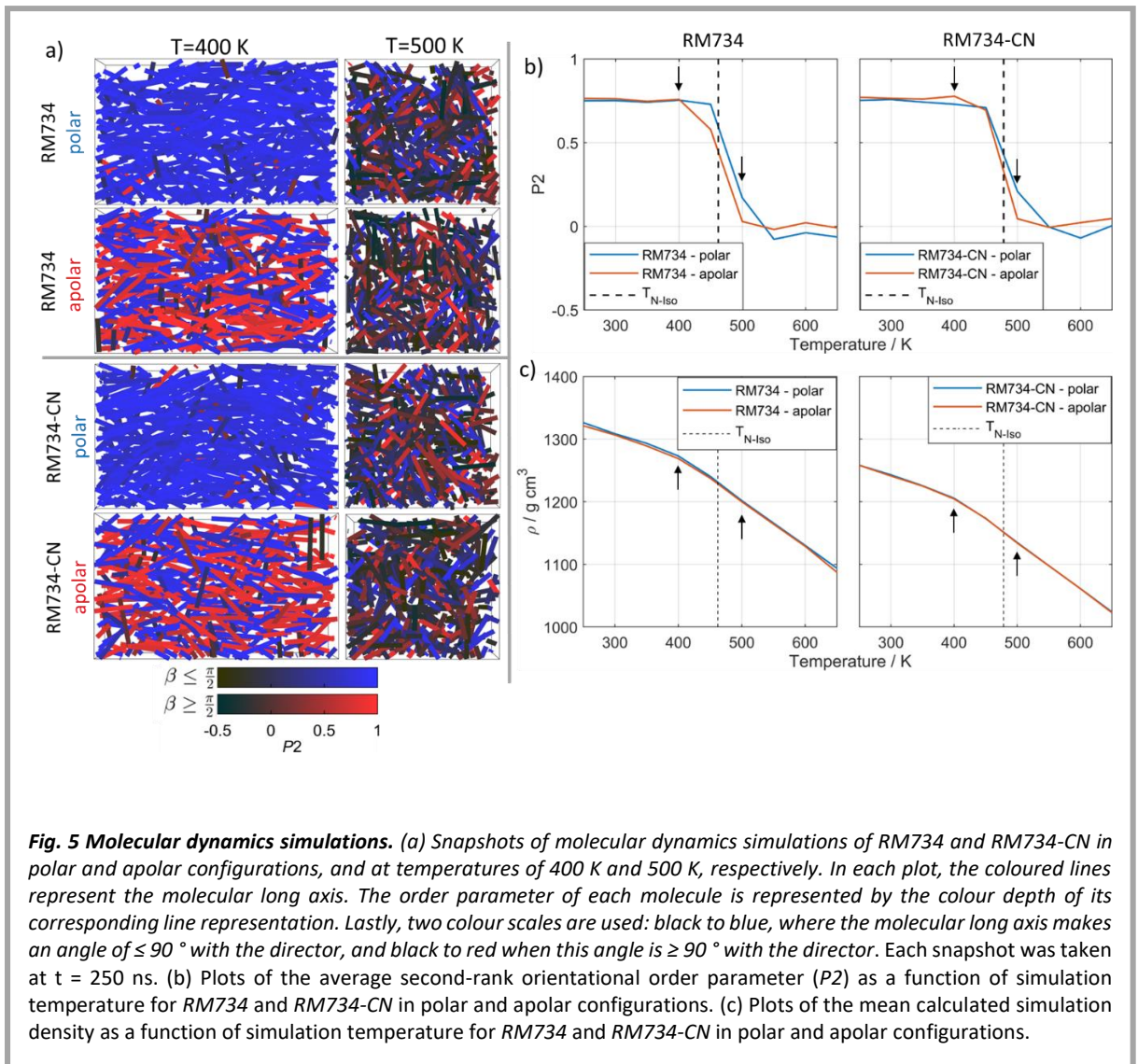

*Fig. 5 Molecular dynamics simulations. (a) Snapshots of molecular dynamics simulations of RM734 and RM734-CN in polar and apolar configurations, and at temperatures of 400 K and 500 K, respectively. In each plot, the coloured lines represent the molecular long axis. The order parameter of each molecule is represented by the colour depth of its corresponding line representation. Lastly, two colour scales are used: black to blue, where the molecular long axis makes an angle of ≤ 90 ° with the director, and black to red when this angle is ≥ 90 ° with the director. Each snapshot was taken at t = 250 ns. (b) Plots of the average second-rank orientational order parameter (P2) as a function of simulation temperature for RM734 and RM734-CN in polar and apolar configurations. (c) Plots of the mean calculated simulation density as a function of simulation temperature for RM734 and RM734-CN in polar and apolar configurations.*

|  | P2 | B | P1(n) | P1 (dipoles) | P (C/m$^2$) | ∠(**n**,**np**)(rad) |
|---|---|---|---|---|---|---|
| *RM734* (POLAR MD) | 0.75 ± 0.005 | 0.036 ± 0.005 | 0.895 ± 0.004 | 0.860 ± 0.008 | 0.064 ± 0.003 | 0.017 ± 0.001 |
| *RM734* (APOLAR MD) | 0.76 ± 0.010 | 0.044 ± 0.011 | 0.017 ± 0.003 | 0.042 ± 0.010 | 0.0017 ± 0.0002 | 0.052 ± 0.003 |
| *RM734-CN* (POLAR MD) | 0.73 ± 0.005 | 0.012 ± 0.006 | 0.897 ± 0.004 | 0.855 ± 0.007 | 0.052 ± 0.003 | 0.004 ± 0.0007 |
| *RM734–CN* (APOLAR MD) | 0.77 ± 0.005 | 0.012 ± 0.006 | 0.009 ± 0.003 | 0.022 ± 0.006 | 0.00061 ± 0.0001 | 0.036 ± 0.003 |
| *RM734* (WAXS) | 0.68 | - | - | - | - | - |
| *RM734-CN* (WAXS) | 0.62 | - | - | - | - | - |

*Table 2:* *Second-rank orientational order parameter (P2), biaxial order parameter (B), polar order parameter (P1), order parameter of the polarization vector (P1(dipoles)), polarization vector (P) and angle between the direction of the polarization vector and the director (∠(**n**,**np**)) at 400K for polar or apolar molecular dynamics simulations; all values are an average over each time step in the production MD run (30 – 280 ns) as described in the text, with plus/minus values corresponding to one standard deviation from the mean. Experimental values P2 obtained by WAXS as described in ref 22.*

Having demonstrated our ability to generate polar and apolar nematic configurations of both *RM734* and *RM734-CN*, and given the distinctive X-Ray scattering behaviour of *RM734*, we next calculated the two dimensional X-ray scattering patterns from our MD simulations, subjecting these to the same analysis as used previously for experimental data. For each simulation, we calculated two-dimensional WAXS patterns as an average of trajectories in the time window 200 – 280 ns, as described in the experimental section (See Note SI.F). We next calculated small angle X-ray scattering intensities as a function of *Q* every 0.5 ns for frames in the time window 200 – 280 ns; values presented in Fig. 6.(e-f) are an average of these 160 frames to remove any effect from instantaneous positional order. Notably X-ray scattering intensities calculated for polar nematic simulations of both *RM734* and *RM734-CN* feature multiple low angle peaks, whereas the corresponding apolar nematic simulations do not. The implication being that the presence of multiple low angle scattering peaks are a consequence of polar nematic order rather than differences in molecular structure. We now compare this with experimental X-ray scattering data (Fig. 6g): *RM734* exhibits several low angle peaks in both the nematic and $N_S$ phases (as do structurally related materials reported previously)[9,10] however, these additional low angle peaks are absent in scattering patterns obtained for the nitrile terminated material *RM734-CN*, which exhibits a typical apolar nematic phase. These results indicate small angle X-ray scattering is perhaps a useful, albeit indirect, probe of polar order within nematic and polar nematic liquid crystals.

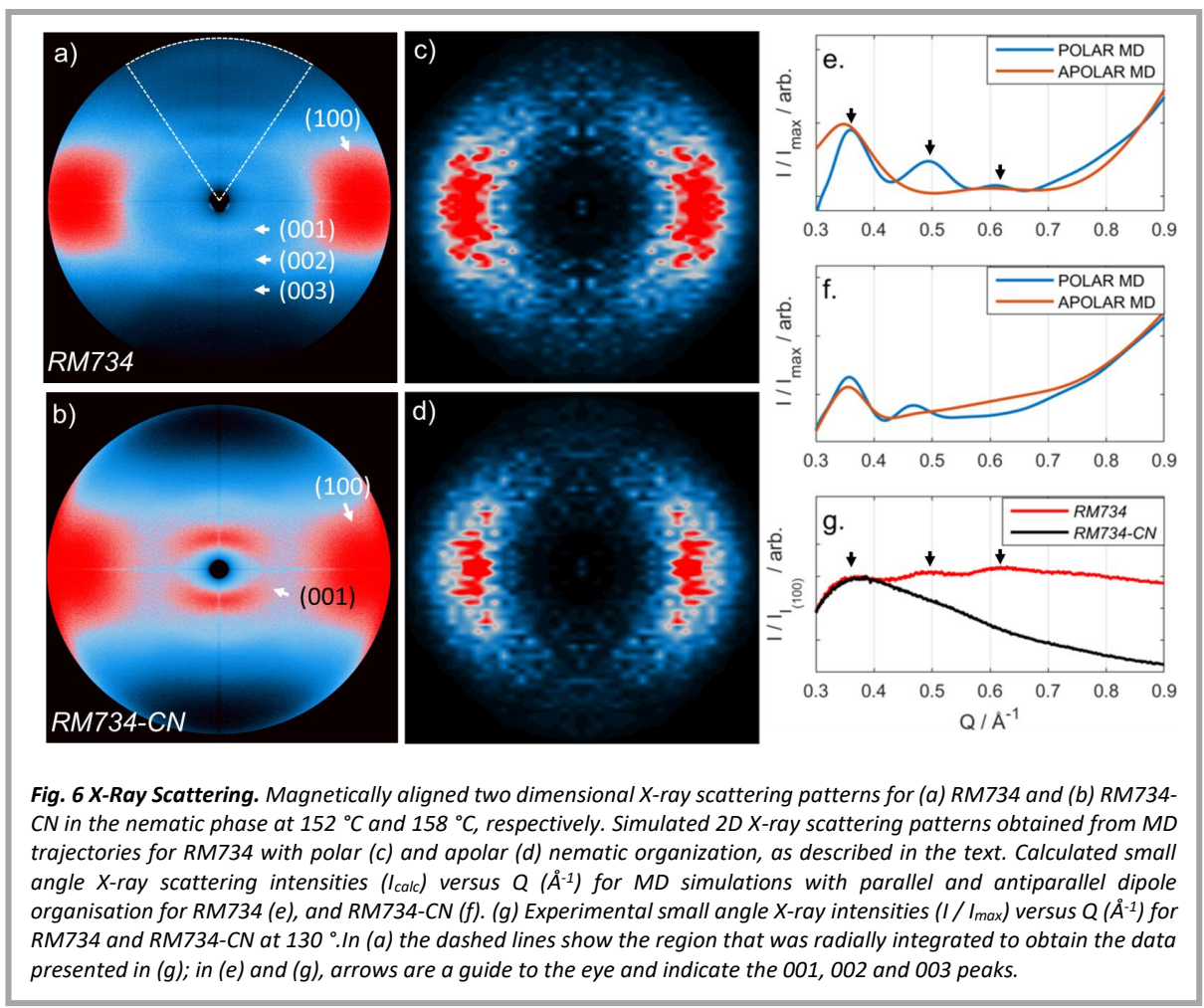

**Fig. 6 X-Ray Scattering.** *Magnetically aligned two dimensional X-ray scattering patterns for (a) RM734 and (b) RM734-CN in the nematic phase at 152 °C and 158 °C, respectively. Simulated 2D X-ray scattering patterns obtained from MD trajectories for RM734 with polar (c) and apolar (d) nematic organization, as described in the text. Calculated small angle X-ray scattering intensities ($I_{calc}$) versus Q (Å$^{-1}$) for MD simulations with parallel and antiparallel dipole organisation for RM734 (e), and RM734-CN (f). (g) Experimental small angle X-ray intensities (I / $I_{max}$) versus Q (Å$^{-1}$) for RM734 and RM734-CN at 130 °. In (a) the dashed lines show the region that was radially integrated to obtain the data presented in (g); in (e) and (g), arrows are a guide to the eye and indicate the 001, 002 and 003 peaks.*

## Discussion

Up to now, nematic phases did not show long-range polar order, even when formed by rod-like molecules with large electric dipole moments. In a uniform nematic phase, random head/tail arrangement is favourable from the entropic point of view as well as from the dipole interaction point of view because antiparallel nearest neighbour´s orientation of elongated dipoles is energetically favourable. In *RM734,* it has been demonstrated that a thermotropic polar nematic phase can be realized in a material made of slightly wedge-shaped molecules. However, the *RM734-CN* analogue reported here shows how a subtle chemical substitution completely modifies this phase behaviour. The first remarkable observation stemming from the comparison performed on both analogues is that, although splay elastic constant is low for *RM734-CN*, no pretransitional softening of it is detected as in the case of *RM734*. Another important observation arises from the evolution of the birefringence Δn vs *P2* (Fig. SI.3). Both materials have comparable polarizabilities, but noticeable different behaviour. The link between birefringence and *P2*, besides depending on the angular distribution function, is also influenced by the orientational molecular correlations[26]. This indicates that orientational molecular correlations differ in both materials. Such fact also relates to the dielectric behaviour of both analogues. While *RM734* shows strong polar correlations culminating in the *N-N$_S$* transition, the onset of such orientational correlations for *RM734-CN*, although discernable, is of much lower magnitude. The opposite difference is observed in terms of positional correlations between nearest neighbours. SAXS measurements evidence weaker positional correlations in the *N* phase of *RM734* than in the case of *RM734-CN*, as inferred from the much weaker scattering intensity when compared to typical nematic scattering patterns.

These great differences between the two very similar analogues raise the question of which are the underlying mechanisms driving the formation of the *N$_S$* phase. As we have shown, taken on their own, electronic structure calculations of isolated molecules, molecular conformational potential energy surfaces, and intermolecular complex formation cannot offer a plausible explanation for the formation of the splay nematic phase and its polar nematic order. On the other hand, the results of molecular dynamics simulations offer several key observations.

The difference in density for the apolar and polar configurations of *RM734*, although small, can only be due to a more efficient packing of the molecules in the polar phase, i.e. polar order of the slightly wedge-shaped *RM734* molecules, reduces the excluded volume, and, simultaneously, decreases attractive interaction energy (Table SI.2). Such a difference is not observed for *RM734-CN*, which only shows the *N* phase. As the dielectric study shows, the polar correlations are also present in this material, but they do not grow when the temperature is decreased. This suggests that the packing efficiency, which results in the reduction of the excluded volume and decrease of the attraction energy, is the underlying mechanism for the appearance of the polar nematic phase. In agreement with these observations, Gregorio *et al* showed that for a simple model of conical molecules, when density increases polar order can build up causing the softening of $K_1$[27]. They state, that in their model, splay elastic constant shows a strong dependence on the molecular geometry. While larger radii gradients of the conical molecules strongly decrease $K_1$, its reduction increases $K_1$ and the usual rod-like behaviour is recovered for uniform radii. Inspired by this hypothesis, we measured the temperature dependence of relative changes of density for *RM734* by monitoring the temperature changes of the length of a slab of the material confined in a sealed capillary (Fig. 7a). Results show that below the *N-N$_S$* transition there is a 0.1-0.2% density change between the extrapolated value for the *N* phase and the measured value for the *N$_S$* phase (Fig. 7b). Such difference is smaller than that obtained from MD simulations (~0.5 %), which is not surprising because polar order gradually grows already in the nematic phase and, at the weakly first order phase transition, only part of the density difference occurs. The question then is why a nitro group results in more efficient packing in the polar configuration than the nitrile analogue. A possible explanation can be found in the calculated molecular pairing configurations (Fig. 4). Although, on their own, such results might not be conclusive, revisited in the current context, the optimized configuration of a pair of *RM734* molecules results in a significantly smaller angle between the molecular axes for parallel pairs (8°) than the antiparallel pairs (15°), while the trend is opposite for *RM734-CN*, for which the angle is 27° for parallel, and 14° for the antiparallel pair (Table 1). In all cases, the complexation is energetically favourable. This comparison also suggests that the attractive interaction between the nitro and the ester (COO) group plays a decisive role in the parallel configuration, as it is absent in the *RM734-CN* pair. Indeed, in the crystal structure of *RM734* we observe a close contact between an oxygen atom ($NO_2$) and carbon atom (COO), with a distance of 3.33 Å (*i.e.* roughly the sum of the VDV radii) (SI.12). The optimized pair configurations also exhibit slightly different conformations, however, the conformers' population analysis of the MD simulation shows that, while there is a difference between the polar and apolar states, the difference between the materials is subtle (Fig.SI.6, Fig.SI.7 and Fig.SI.8) The large angle for the *RM734-CN* parallel pair is also consistent with notably smaller $P2$ observed in the polar MD simulation of this material (Table 2). With this in mind, we calculated the optimized configuration pairs for *RM554*, a previously reported analogue of *RM734* where a fluorine atom is positioned ortho to the nitro group and the terminal chain is lengthened by one carbon group (Figure SI.11)[9]. We note that *RM554* exhibits $N_S$ phase, and as in the case of *RM734*, optimised configuration of a pair of molecules show that in the parallel configuration the molecular axes of the two molecules are more aligned (relative angle 8°) than in the antiparallel case (13°) (Fig. SI. 11 and Table SI.3). Staggering of the parallel pairs is also observed, with the nitro and ester groups coming together. This material shows the same distinct X-ray scattering patterns as *RM734*. Altogether, these observations highlight the possibility of using MD simulations as a predictive tool for molecular design; in this cases by comparing densities of polar and non-polar configurations, but also through computation of other properties, in the pursuit of materials showing the polar nematic phase in the desired temperature range. As a predictor of the efficiency of packing, promising materials could be identified by comparison of the complexation geometries of parallel and antiparallel pairs from DFT calculations at a lower computational expense than MD simulations, which would then be performed for the most interesting of materials highlighted by this method.

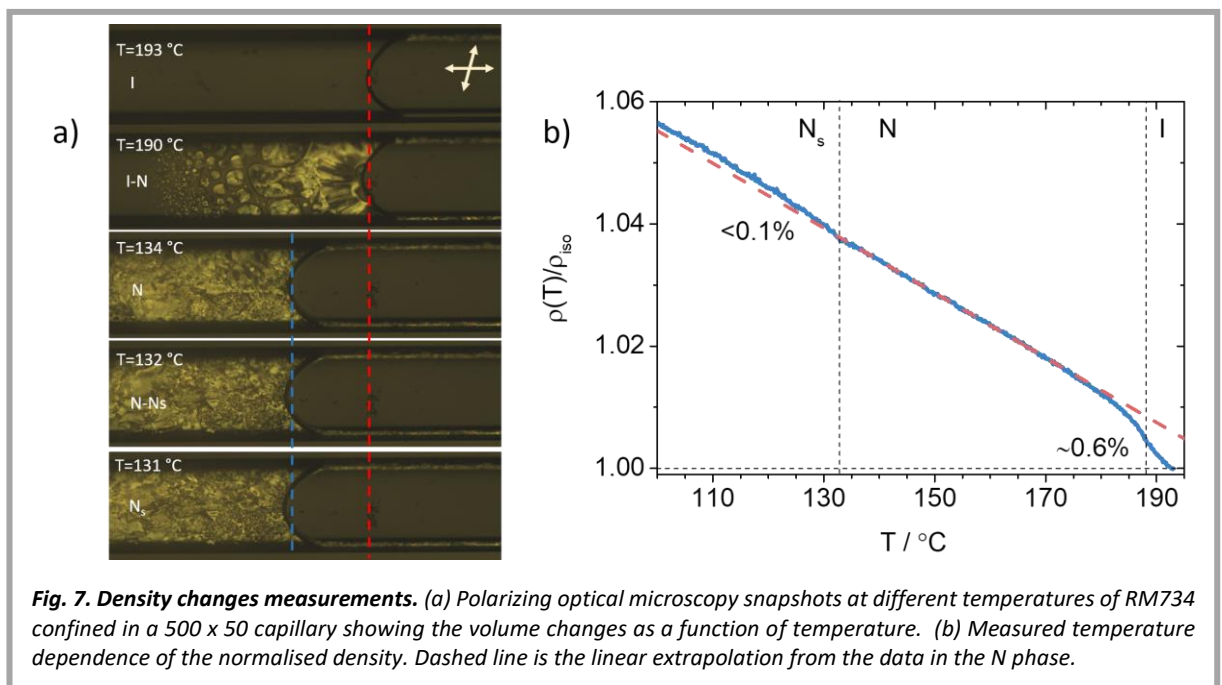

***Fig. 7. Density changes measurements.*** *(a) Polarizing optical microscopy snapshots at different temperatures of RM734 confined in a 500 x 50 capillary showing the volume changes as a function of temperature. (b) Measured temperature dependence of the normalised density. Dashed line is the linear extrapolation from the data in the N phase.*

We could further link MD simulations results to our experimental data. The additional small angle X-ray scattering peaks observed for splay-nematic materials such as *RM734* appear to be a consequence of the unique dipole ordering exhibited by these materials which, inter alia, leads to the formation of the $N_S$ phase and its remarkable properties. MD simulations of the polar state can faithfully reproduce these additional low angle reflections, whereas apolar nematic simulations cannot. The staggered parallel pairing suggested by DFT calculations is also observed in the bulk phase during molecular dynamics simulations of the polar nematic configuration of *RM734* (Fig.SI.9). Although such molecular shift seems to be the origin of the additional SAXS peaks, the small positional correlations evidenced by the weak scattering signal suggest that some staggered positions are only slightly more favourable than others. Existence of the characteristic weak X-ray scattering pattern and the small angles peaks in *RM734* already in the nematic phase far away from the phase transition to the ferroelectric splay nematic shows that local polar order is dominant already at high temperatures. The exceptional agreement between experimental and simulated X-ray scattering data shows that weak positional correlations and the appearance of additional peaks in the direction along the director may be used to predict and differentiate $N_S$ and precursor *N* phases from classical nematics.

Ferroelectric nematics have huge potential for applications. However, finding such materials has not been a trivial task as demonstrated by the long time it has taken to realize this new phase of matter despite the efforts. Pairing optimized geometries and MD simulations of density and X-ray patterns, each on their own, might not be conclusive. However, altogether the knowledge we gain from MD simulations suggests that ferroelectric (splay) nematics could be designed *in silico*, rather than being discovered on an *ad hoc* basis, with the potential to greatly accelerate applications utilizing this new molecular organization.

**Materials and Methods**

**Materials**

Both *RM734* and *RM734-CN* were synthesised via literature methods[9]; their chemical structures and transition temperatures (°C) are given in Fig. 1, with monotropic phase transitions presented in parenthesis ()[9].

**X-ray Scattering**

The X-ray scattering setup is described elsewhere, and datasets are reported elsewhere[9]. Values of Q were calibrated against a standard of silver behenate. The scattering pattern from an empty capillary (under the same experimental conditions as for each sample) was used as a background and was subtracted from each frame prior to analysis.

**Quantum Chemical Calculations**

Computational chemistry was performed in Gaussian G09 rev D01[28] on either the ARC3 machine at the University of Leeds, or using the same software package on the YARCC or Viking machines at the University of York. Output files were rendered using Qutemol,[29] or VMD [30]. Calculations utilised the M06-HF[31] hybrid DFT functional with additional D3 dispersion correction,[32] the aug-cc-pVTZ basis set[33], counterpoise correction to basis set superposition error (BSSE)[34]. The keywords Integral=UltraFine and SCF(maxcycles=1024) were used to ensure convergence, while a frequency calculation was used to confirm the absence of imaginary frequencies and so confirm the optimised geometries were true minima.

Fully relaxed scans along the potential energy surface for a given dihedral were performed at the M06HF-D3/aug-cc-pVTZ level of DFT with a step size of 12 ° via the OPT=MODREDUNDANT keyword.

Polarizabilities and hyperpolarizabilities for *RM734* and *RM734-CN* were calculated at the M06HF-D3/aug-cc-pVTZ level of DFT *via* the POLAR keyword, at frequencies of 400 nm and 800 nm (specified *via* CPHF=RdFreq).

For calculation of interaction energies, geometry optimisation was performed on homogenous pairs *RM734* and *RM734-CN* in parallel and anti-parallel orientations. No geometry constraints were employed during optimisation beyond the initial starting configuration.

**Molecular Dynamics**

Fully atomistic molecular dynamics (MD) simulations were performed either in Gromacs 2019.3 [35–41] on the ARC4 machine at the University of Leeds, with support for NVIDIA V100 GPUs through CUDA 10.1.168, or in Gromacs 2016.2 on the YARCC HPC at the University of York  We used the General Amber Force Field[42] with modifications for liquid crystalline molecules[43]. Topologies were generated using AmberTools 16[44,45] and converted into Gromacs readable format with Acpype[46]. Atomic charges were determined using the RESP method[47] for geometries optimised at the B3LYP/6-31G(d) level of DFT[48,49] using the Gaussian G09 revision d01 software package[28].

We constructed initial low density lattices of 680 molecules with random positional order. For polar nematic simulations, all molecular dipole moments were oriented along +x, while apolar nematic simulations have a 1:1 mixture of molecules aligned along +x and –x. Following energy minimisation, the simulation box was then compressed over 50 ps to a mass density of ~ 1 g cm$^3$ which is typical of that of low molecular weight liquid crystals. Following compression, simulations were allowed to equilibrate for 30 ns, with subsequent production MD runs of a further 250 ns. Temperature variations were studied by taking the final topology and trajectory of a completed simulation at 400 K, and heating (or cooling) to the desired temperature by setting *ref-t* to an appropriate value; we then performed a 30 ns equilibration followed by a further 250 ns production MD run. Analysis was performed during the time period 30 – 280 ns, i.e. the production MD run only. A timestep of 0.5 fs was used, and trajectories were recorded every 10 ps.

Simulations employed periodic boundary conditions in xyz. Bonds lengths were constrained to their equilibrium values with the LINCS algorithm [50]. System pressure was maintained at 1 bar using anisotropic Parrinello-Rahamn pressure coupling[51,52], enabling the relative box dimensions to independently vary in all dimensions. Compressabilities in xyz dimensions were set to 4.5e-5, with the off-diagonal compressibilities were set to zero to ensure the simulation box remained rectangular. Simulation temperature was controlled with a Nosé–Hoover thermostat[53,54]. Long-range electrostatic interactions were calculated using the Particle Mesh Ewald method with a cut-off value of 1.2 nm[55]. A van der Waals cut-off of 1.2 nm was used. MD trajectories were visualised using PyMOL 4.5.

High resolution one dimensional SAXS curves (I versus Q) were computed from MD trajectories using CRYSOL[56]. Simulated two dimensional WAXS patterns were computed from MD trajectories using a modification of the procedure described by Coscia et al.[57]. Calculation of orientational order parameters from azimuthally integrated simulated 2D WAXS patterns used the method described in refs[58,59]

**Broadband dielectric spectroscopy**

Spectroscopy measurements of the complex dielectric permittivity $\varepsilon(\omega) = \varepsilon'(\omega) - i\varepsilon''(\omega)$ were carried out in the $10^3 - 1.1 \cdot 10^8$ Hz frequency range with the HP4294 ($10^3$-$1.1 \cdot 10^8$ Hz) impedance analyzers. The material was placed in the nematic phase between two circular gold-plated brass electrodes (5 mm diameter) acting as a parallel-plate capacitor. The separation between electrodes was fixed by 50 μm thick silica spacers. This sample was accommodated in a modified HP16091A coaxial fixture with a sliding short-circuit along its 7 mm coaxial line section and central conductor. Calibration of the devices allowed for excellent overlap of the measured spectra. The temperature of the sample was controlled down to ±50 mK in a Novocontrol cryostat. From the comparison with results obtained for the electrodes treated for homeotropic alignment, those obtained in classical glass ITO cells for parallel alignment, and observations of the orientation of the *N* phase when in direct contact with the ITO surface, we concluded that in the bare gold electrodes the sample spontaneously aligns homeotropically in the nematic phase. All measurements were performed on cooling from the isotropic phase.

**Dynamic light scattering**

In the DLS experiments, we used a standard setup, using a frequency-doubled diode-pumped ND:YAG laser (532 nm, 80 mW), an ALV APD based "pseudo" cross-correlation detector, and ALV-6010/160 correlator to obtain the autocorrelation function of the scattered light intensity. The direction and the polarization of the incoming and detected light were chosen so that pure splay mode was observed [13]. A single mode optical fibre with a GRIN lens was used to collect the scattered light within one coherence area. We fitted the intensity autocorrelation function $g_2$ with $g_2 = 1 + 2(1-j_d)j_d g_1 + j_d^2 g_1^2$, where $j_d$ is the ratio between the intensity of the light that is scattered inelastically and the total scattered intensity, and $g_1$ was a single exponential function, $g_1 = \mathrm{Exp}(-t/\tau)$ The relaxation rate $1/\tau$ was attributed to the splay eigenmode of orientational fluctuations with the wavevector q equal to the scattering vector $q_s$. The scattered intensity of the mode was determined as a product $j_d I_{tot}$, where $I_{tot}$ was the total detected intensity. The scattered intensity from a pure splay mode is, $I_1 \propto (\Delta\varepsilon_{opt})^2 / K_1 q^2$, and the relaxation rates $1/\tau_1 = K_1 q^2 / \eta_1$, where $q$ is the scattering vector[26]. The temperature dependence of the anisotropy of dielectric tensor at optical frequencies $\Delta\varepsilon_{opt}$ was obtained from measurements of Δ*n* (Fig. SI.3). With this method, the temperature dependence of the splay elastic constant and viscosity are obtained, but not their absolute value. We determined the absolute values of $K_1$ and $K_3$ at 180 °C from capacitance measurement of the Fredericks transition, where the threshold voltage for the reorientation of the material in a planar cell is related to the splay elastic constant by the relation $V_{th} = \sqrt{\pi^2 K_1 / \varepsilon_0 \Delta\varepsilon}$ (See SI Note C).

**Normalised density changes**

Changes in density where measured by polarizing optical microscopy. A section of a capillary of 500 μm width and 50 μm was filled with the material and sealed at both ends. The capillary was placed between slightly uncrossed polarizers for better definition of the meniscus position. The sample was heated to the isotropic phase and then cooled at 2 °C/min while recording. Temperature gradients were avoided with a copper enclosure for the capillary. Changes in density are then calculated from the variation of the total material slab length referenced to the isotropic phase, $\rho(T)/\rho_{iso} = V_{iso}/V(T) = L_{iso}/L(T)$

## Acknowledgements

Computational work was undertaken on either ARC3 or ARC4, part of the High Performance

Computing facilities at the University of Leeds, UK, or the YARCC or Viking High Performance Computing facilities at the University of York. N. S. and A. M. acknowledge the financial support from the Slovenian Research Agency (research core Funding No. P1-0192). J.M-P acknowledges funding by the University of the Basque Country (project GIU18/146). N.S and J.M-P thank Prof. M.R. de la Fuente for technical support with the dielectric measurements.

## Author contribution

R.J.M synthesised the materials, performed X-ray experiments, DFT and MD simulations. N.S and J.M-P performed dielectric spectroscopy measurements and contributed to its analysis. N.S and A.M. carried out DLS experiments. N.S, J.M-P and A.M collected birefringence data. N.S and A.M coordinated the work. A.M oversaw all the contributions. R.J.M, N.S and A.M prepared the initial draft of the manuscript and all the authors made contributions to the final version.

## Competing interests

The authors declare no competing interests.

## Additional information

Supplementary information is available for this paper

## References

1. Mertelj, A., Lisjak, D., Drofenik, M. & Copic, M. Ferromagnetism in suspensions of magnetic platelets in liquid crystal. *Nature* **504**, 237–241 (2013).
2. Shuai, M. *et al.* Spontaneous liquid crystal and ferromagnetic ordering of colloidal magnetic nanoplates. *Nat Commun* **7**, 10394 (2016).
3. Sebastián, N. *et al.* Ferroelectric-Ferroelastic Phase Transition in a Nematic Liquid Crystal. *Phys. Rev. Lett.* **124**, 037801 (2020).
4. Morozov, K. I. Long-range order of dipolar fluids. *The Journal of Chemical Physics* **119**, 13024–13032 (2003).
5. Palffy-Muhoray, P., Lee, M. A. & Petschek, R. G. Ferroelectric Nematic Liquid Crystals: Realizability and Molecular Constraints. *Phys. Rev. Lett.* **60**, 2303–2306 (1988).
6. Liu, Q., Ackerman, P. J., Lubensky, T. C. & Smalyukh, I. I. Biaxial ferromagnetic liquid crystal colloids. *Proceedings of the National Academy of Sciences* 201601235 (2016) doi:10.1073/pnas.1601235113.
7. Ackerman, P. J. & Smalyukh, I. I. Static three-dimensional topological solitons in fluid chiral ferromagnets and colloids. *Nat Mater* **16**, 426–432 (2017).
8. Rupnik, P. M., Lisjak, D., Čopič, M., Čopar, S. & Mertelj, A. Field-controlled structures in ferromagnetic cholesteric liquid crystals. *Science Advances* **3**, e1701336 (2017).
9. Mandle, R. J., Cowling, S. J. & Goodby, J. W. Rational Design of Rod-Like Liquid Crystals Exhibiting Two Nematic Phases. *Chem. Eur. J.* **23**, 14554–14562 (2017).
10. J. Mandle, R., J. Cowling, S. & W. Goodby, J. A nematic to nematic transformation exhibited by a rod-like liquid crystal. *Physical Chemistry Chemical Physics* **19**, 11429–11435 (2017).
11. Connor, P. L. M. & Mandle, R. J. Chemically induced splay nematic phase with micron scale periodicity. *Soft Matter* **16**, 324–329 (2020).
12. Sebastián, N., Mandle, R. J., Petelin, A., Eremin, A. & Mertelj, A. Electric switching in a polar nematic phase. *to be submitted* (2020).
13. Mertelj, A. *et al.* Splay Nematic Phase. *Phys. Rev. X* **8**, 041025 (2018).
14. Berardi, R., Ricci, M. & Zannoni, C. Ferroelectric and Structured Phases from Polar Tapered Mesogens. *Ferroelectrics* **309**, 3–13 (2004).

15. Rosseto, M. P. & Selinger, J. V. Theory of the splay nematic phase: Single versus double splay. *Phys. Rev. E* **101**, 052707 (2020).
16. Pleiner, H. & Brand, H. R. Spontaneous Splay Phases in Polar Nematic Liquid Crystals. *EPL* **9**, 243 (1989).
17. Chaturvedi, N. & Kamien, R. D. Gnomonious projections for bend-free textures: thoughts on the splay-twist phase. *Proceedings of the Royal Society A: Mathematical, Physical and Engineering Sciences* **476**, 20190824 (2020).
18. Chen, X. *et al.* First-principles experimental demonstration of ferroelectricity in a thermotropic nematic liquid crystal: Polar domains and striking electro-optics. *Proc Natl Acad Sci USA* **117**, 14021 (2020).
19. Nishikawa, H. *et al.* A Fluid Liquid-Crystal Material with Highly Polar Order. *Advanced Materials* **29**, 1702354 (2017).
20. Havriliak, S. & Negami, S. A Complex Plane Analysis of alfa-Dispersions in Some Polymer Systems. *J. Polym. Sci. C* **14**, 99–117 (1966).
21. Luigi Nordio, P., Rigatti, G. & Segre, U. Dielectric relaxation theory in nematic liquids. *Molecular Physics* **25**, 129–136 (1973).
22. Mandle, R. J. & Mertelj, A. Orientational order in the splay nematic ground state. *Phys. Chem. Chem. Phys.* **21**, 18769–18772 (2019).
23. Peláez, J. & Wilson, M. Molecular orientational and dipolar correlation in the liquid crystal mixture E7: a molecular dynamics simulation study at a fully atomistic level. *Phys. Chem. Chem. Phys.* **9**, 2968–2975 (2007).
24. Tiberio, G., Muccioli, L., Berardi, R. & Zannoni, C. Towards in Silico Liquid Crystals. Realistic Transition Temperatures and Physical Properties for n-Cyanobiphenyls via Molecular Dynamics Simulations. *ChemPhysChem* **10**, 125–136 (2009).
25. Sims, M. T., Abbott, L. C., Cowling, S. J., Goodby, J. W. & Moore, J. N. Dyes in Liquid Crystals: Experimental and Computational Studies of a Guest–Host System Based on a Combined DFT and MD Approach. *Chemistry – A European Journal* **21**, 10123–10130 (2015).
26. Gennes, P. G. de & Prost, J. *The Physics of Liquid Crystals*. (Clarendon Press, 1995).
27. Gregorio, P. D., Frezza, E., Greco, C. & Ferrarini, A. Density functional theory of nematic elasticity: softening from the polar order. *Soft Matter* **12**, 5188–5198 (2016).
28. M. J. Frisch, G. W. Trucks, H. B. Schlegel, G. E. Scuseria, M. A. Robb, J. R. Cheeseman, G. Scalmani, V. Barone, G. A. Petersson, H. Nakatsuji, X. Li, M. Caricato, A. Marenich, J. Bloino, B. G. Janesko, R. Gomperts, B. Mennucci, H. P. Hratchian, J. V. Ortiz, A. F. Izmaylov, J. L. Sonnenberg, D. Williams-Young, F. Ding, F. Lipparini, F. Egidi, J. Goings, B. Peng, A. Petrone, T. Henderson, D. Ranasinghe, V. G. Zakrzewski, J. Gao, N. Rega, G. Zheng, W. Liang, M. Hada, M. Ehara, K. Toyota, R. Fukuda, J. Hasegawa, M. Ishida, T. Nakajima, Y. Honda, O. Kitao, H. Nakai, T. Vreven, K. Throssell, J. A. Montgomery, Jr., J. E. Peralta, F. Ogliaro, M. Bearpark, J. J. Heyd, E. Brothers, K. N. Kudin, V. N. Staroverov, T. Keith, R. Kobayashi, J. Normand, K. Raghavachari, A. Rendell, J. C. Burant, S. S. Iyengar, J. Tomasi, M. Cossi, J. M. Millam, M. Klene, C. Adamo, R. Cammi, J. W. Ochterski, R. L. Martin, K. Morokuma, O. Farkas, J. B. Foresman, and D. J. Fox. *Gaussian 09, Revision A.02*. (Gaussian, Inc., 2016).
29. Tarini, M., Cignoni, P. & Montani, C. Ambient Occlusion and Edge Cueing for Enhancing Real Time Molecular Visualization. *IEEE Transactions on Visualization and Computer Graphics* **12**, 1237–1244 (2006).
30. Humphrey, W., Dalke, A. & Schulten, K. VMD: Visual molecular dynamics. *Journal of Molecular Graphics* **14**, 33–38 (1996).
31. Zhao, Y. & Truhlar, D. G. Density Functional for Spectroscopy: No Long-Range Self-Interaction Error, Good Performance for Rydberg and Charge-Transfer States, and Better Performance on Average than B3LYP for Ground States. *J. Phys. Chem. A* **110**, 13126–13130 (2006).
32. Grimme, S., Antony, J., Ehrlich, S. & Krieg, H. A consistent and accurate ab initio parametrization of density functional dispersion correction (DFT-D) for the 94 elements H-Pu. *J. Chem. Phys.* **132**, 154104 (2010).
33. Kendall, R. A., Dunning, T. H. & Harrison, R. J. Electron affinities of the first-row atoms revisited. Systematic basis sets and wave functions. *J. Chem. Phys.* **96**, 6796–6806 (1992).
34. Boys, S. F. & Bernardi, F. The calculation of small molecular interactions by the differences of separate total energies. Some procedures with reduced errors. *Molecular Physics* **19**, 553–566 (1970).
35. Berendsen, H. J. C., van der Spoel, D. & van Drunen, R. GROMACS: A message-passing parallel molecular dynamics implementation. *Computer Physics Communications* **91**, 43–56 (1995).
36. Lindahl, E., Hess, B. & van der Spoel, D. GROMACS 3.0: a package for molecular simulation and trajectory analysis. *J Mol Model* **7**, 306–317 (2001).
37. Van Der Spoel, D. *et al.* GROMACS: fast, flexible, and free. *J Comput Chem* **26**, 1701–1718 (2005).

38. Hess, B., Kutzner, C., van der Spoel, D. & Lindahl, E. GROMACS 4: Algorithms for Highly Efficient, Load-Balanced, and Scalable Molecular Simulation. *J. Chem. Theory Comput.* **4**, 435–447 (2008).
39. Pronk, S. *et al.* GROMACS 4.5: a high-throughput and highly parallel open source molecular simulation toolkit. *Bioinformatics* **29**, 845–854 (2013).
40. Abraham, M. J. *et al.* GROMACS: High performance molecular simulations through multi-level parallelism from laptops to supercomputers. *SoftwareX* **1–2**, 19–25 (2015).
41. Páll, S., Abraham, M. J., Kutzner, C., Hess, B. & Lindahl, E. Tackling Exascale Software Challenges in Molecular Dynamics Simulations with GROMACS. in *Solving Software Challenges for Exascale* (eds. Markidis, S. & Laure, E.) vol. 8759 3–27 (Springer International Publishing, 2015).
42. Wang, J., Wolf, R. M., Caldwell, J. W., Kollman, P. A. & Case, D. A. Development and testing of a general amber force field. *Journal of Computational Chemistry* **25**, 1157–1174 (2004).
43. Boyd, N. J. & Wilson, M. R. Optimization of the GAFF force field to describe liquid crystal molecules: the path to a dramatic improvement in transition temperature predictions. *Phys. Chem. Chem. Phys.* **17**, 24851–24865 (2015).
44. Case, D. A. *et al.* The Amber biomolecular simulation programs. *Journal of Computational Chemistry* **26**, 1668–1688 (2005).
45. Wang, J., Wang, W., Kollman, P. A. & Case, D. A. Automatic atom type and bond type perception in molecular mechanical calculations. *Journal of Molecular Graphics and Modelling* **25**, 247–260 (2006).
46. Sousa da Silva, A. W. & Vranken, W. F. ACPYPE - AnteChamber PYthon Parser interfacE. *BMC Research Notes* **5**, 367 (2012).
47. Bayly, C. I., Cieplak, P., Cornell, W. & Kollman, P. A. A well-behaved electrostatic potential based method using charge restraints for deriving atomic charges: the RESP model. *J. Phys. Chem.* **97**, 10269–10280 (1993).
48. Becke, A. D. Density-functional thermochemistry. III. The role of exact exchange. *J. Chem. Phys.* **98**, 5648–5652 (1993).
49. Lee, C., Yang, W. & Parr, R. G. Development of the Colle-Salvetti correlation-energy formula into a functional of the electron density. *Phys. Rev. B* **37**, 785–789 (1988).
50. Hess, B., Bekker, H., Berendsen, H. J. C. & Fraaije, J. G. E. M. LINCS: A linear constraint solver for molecular simulations. *Journal of Computational Chemistry* **18**, 1463–1472 (1997).
51. Parrinello, M. & Rahman, A. Polymorphic transitions in single crystals: A new molecular dynamics method. *Journal of Applied Physics* **52**, 7182–7190 (1981).
52. Nosé, S. & Klein, M. L. Constant pressure molecular dynamics for molecular systems. *Molecular Physics* **50**, 1055–1076 (1983).
53. Nosé, S. A molecular dynamics method for simulations in the canonical ensemble. *Molecular Physics* **52**, 255–268 (1984).
54. Hoover, W. G. Canonical dynamics: Equilibrium phase-space distributions. *Phys. Rev. A* **31**, 1695–1697 (1985).
55. Darden, T., York, D. & Pedersen, L. Particle mesh Ewald: An N·log(N) method for Ewald sums in large systems. *J. Chem. Phys.* **98**, 10089–10092 (1993).
56. Svergun, D., Barberato, C. & Koch, M. H. J. CRYSOL – a Program to Evaluate X-ray Solution Scattering of Biological Macromolecules from Atomic Coordinates. *J Appl Cryst* **28**, 768–773 (1995).
57. Coscia, B. J. *et al.* Understanding the Nanoscale Structure of Inverted Hexagonal Phase Lyotropic Liquid Crystal Polymer Membranes. *J. Phys. Chem. B* **123**, 289–309 (2019).
58. Sims, M. T., Abbott, L. C., Richardson, R. M., Goodby, J. W. & Moore, J. N. Considerations in the determination of orientational order parameters from X-ray scattering experiments. *Liquid Crystals* **46**, 11–24 (2019).
59. Agra-Kooijman, D. M., Fisch, M. R. & Kumar, S. The integrals determining orientational order in liquid crystals by x-ray diffraction revisited. *Liquid Crystals* **45**, 680–686 (2018).

# Supporting Information

# On the molecular origins of the ferroelectric splay nematic phase

Richard J. Mandle[1,2], Nerea Sebastián[3], Josu Martinez-Perdiguero[4] & Alenka Mertelj[3]

[1] School of Physics and Astronomy, University of Leeds, Leeds, UK, LS2 9JT
[2] Department of Chemistry, University of York, York, YO10 5DD, UK
[3] Jožef Stefan Institute, SI-1000 Ljubljana, Slovenia
[4] Department of Physics, University of the Basque Country (UPV/EHU), Apdo.644-48080 Bilbao, Spain

**Table of contents**

## A. Dielectric Spectroscopy RM734-CN

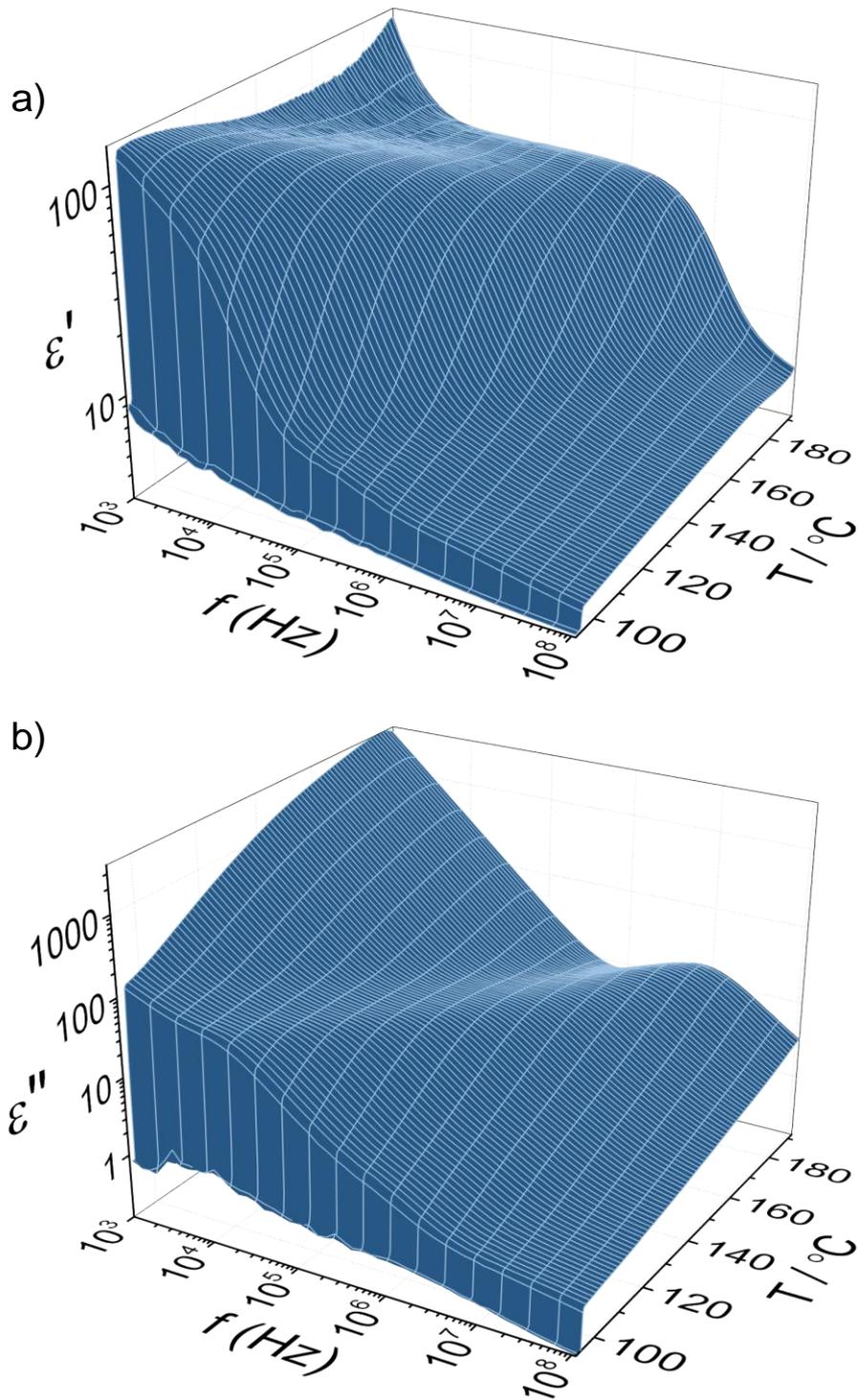

**Figure SI.1: Dielectric spectra of *RM734-CN* in the *N* phase.** Temperature and frequency dependence of the (a) real and (b) imaginary components of the dielectric spectra of *RM734-CN*.

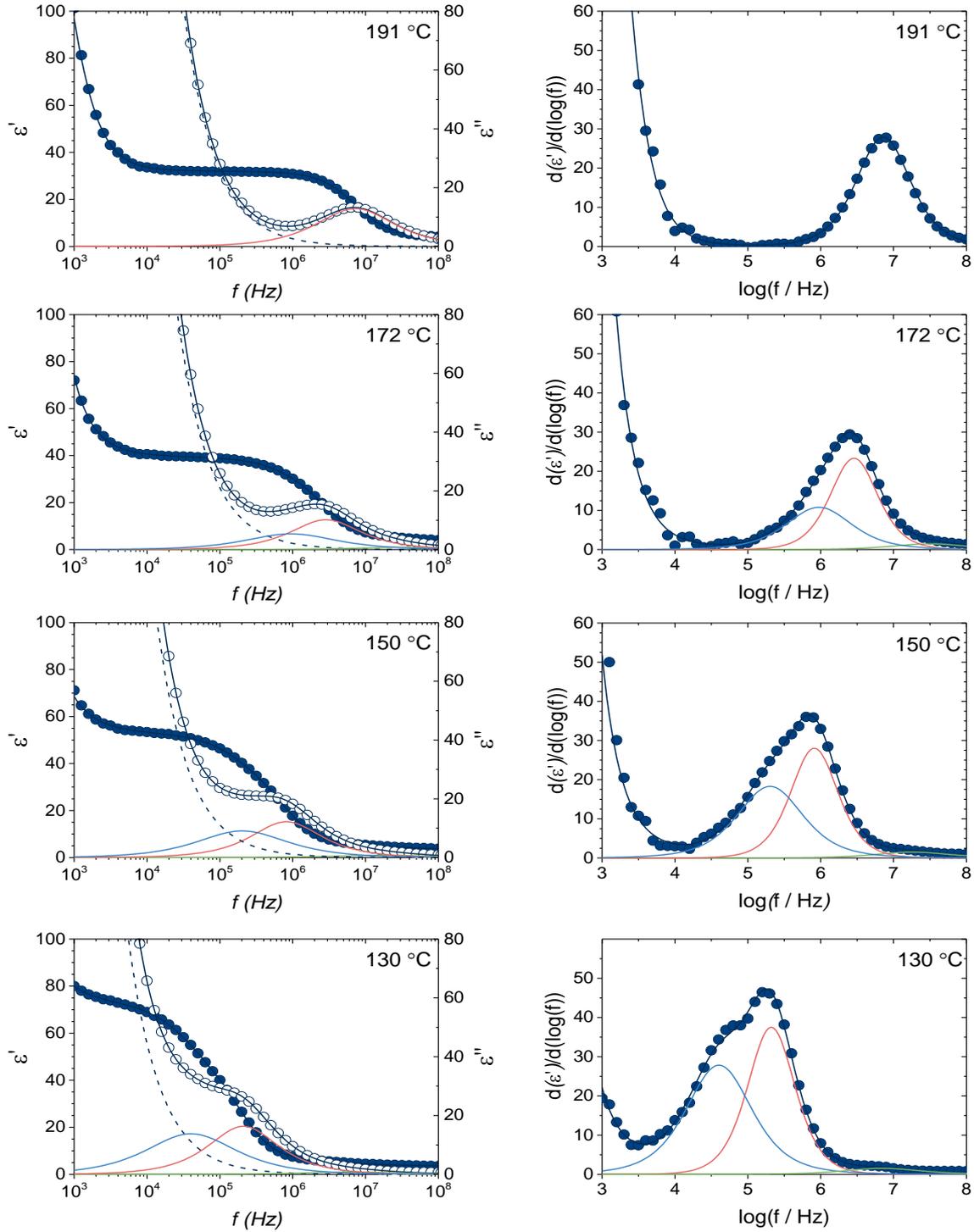

**Figure SI.2:** Examples at different temperatures of the measured dielectric spectra and their fits for ***RM734-CN***. (left) Frequency dependence of the real (squares) and imaginary (circles) dielectric permittivity. Solid lines result from fitting to Equation 1 in the manuscript and the corresponding deconvolution into the elementary processes. Dashed lines correspond to the current conductivity term. (right) The derivative of the real part of the permittivity $d(\varepsilon')/d\log(f)$ at the corresponding temperatures allows for better visualization of the relaxation modes and benevolence of the fits.

## B. Birefringence and order parameter

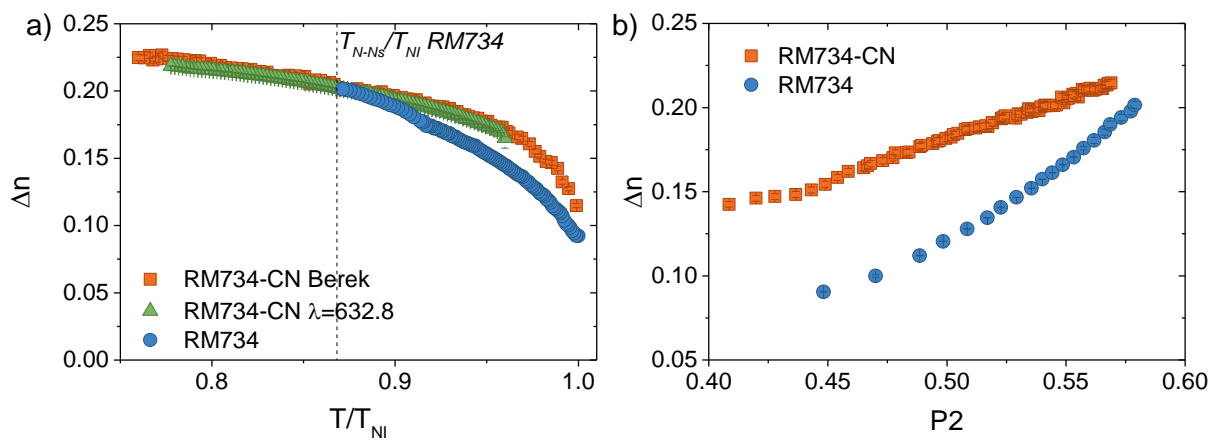

**Figure SI.3**: **Birefringence measurements**. a) Comparison of the temperature dependence of the birefringence of *RM734-CN* and *RM734*. The anisotropy of the index of refraction $\Delta n$ of *RM734-CN* was measured by polarization microscopy. A d=20 µm cell with planar alignment (director in the cell plane) was placed between crossed polarizers with the director at 45 degrees with respect to them. The phase difference between the ordinary and the extraordinary light $\phi = 2\pi\Delta n d/\lambda$ was calculated from the intensity of monochromatic light ($\lambda = 632.8$ nm) transmitted through the sample. Additionally, the temperature dependence of $\Delta n$ was measured with a Berek compensator in a 9 µm cell. As shown by the figure both results are comparable. Results for *RM734-CN* are also compared to $\Delta n$ of *RM734* (reference 13 main manuscript). (b) Representation of the birefringence vs experimental values of *P2* as given in reference 22 in the main manuscript. Given the comparable polarizabilities of both materials, the plot reflects the difference in molecular orientational correlations in the *N* phase of *RM734* and *RM734-CN*.

## C. Fredericks transition

A reference value for the splay elastic constant was measured from the change in the dielectric permittivity when a variable voltage is applied to a planar aligned sample. The frequency was set to 30 kHz to avoid undesired ionic effects and the cell's ITO relaxation. Experimental results were fitted to equations:

$$V = \frac{2V_{th}}{\pi}\sqrt{1+\gamma\eta}\int_{\psi_0}^{\pi/2}\left[\frac{1+\kappa\eta\sin^2\psi}{(1+\gamma\eta\sin^2\psi)(1-\eta\sin^2\psi)}\right]^{1/2}\mathrm{d}\psi \quad \text{(Eq.SI.1)}$$

$$C = C_\perp \frac{\int_{\psi_0}^{\pi/2}\left[\frac{(1+\gamma\eta\sin^2\psi)(1+\kappa\eta\sin^2\psi)}{(1-\eta\sin^2\psi)}\right]^{1/2}\mathrm{d}\psi}{\int_{\psi_0}^{\pi/2}\left[\frac{1+\kappa\eta\sin^2\psi}{(1+\gamma\eta\sin^2\psi)(1-\eta\sin^2\psi)}\right]^{1/2}\mathrm{d}\psi} \quad \text{(Eq.SI.2)}$$

Where the parameter $\eta$ is related to the maximum tilt angle at the centre of the cell $\phi_m$ ($\eta = \sin^2(\phi_m)$); the parameters $\gamma$ and $\kappa$ correspond to the reduced quantities $\gamma = \varepsilon_\perp / \varepsilon_\parallel - 1$ and $\kappa = K_3 / K_1 - 1$. $\varepsilon_\perp$ corresponds to the value of the permittivity below the threshold voltage ($V_{th}$) for the Fredericksz transition. $\varepsilon_\parallel$ corresponds to the dielectric permittivity values for saturated director reorientation. Values of $V_{th}$ and $\Delta\varepsilon$ are then used to calculate the splay elastic constant $K_1 = (V_{th}/\pi)^2 \varepsilon_0 \Delta\varepsilon$

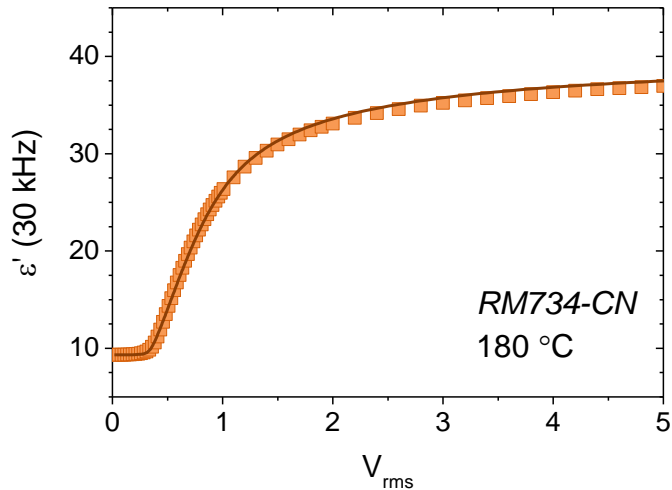

**Figure SI.4: Fredericks transition.** Voltage dependence of the real permittivity at 30 kHz at 180 °C. Solid line shows the fit to Equations SI.1 and SI.2

## D. Polarizabilities, inertia tensor and dipole direction of *RM734* and *RM734-CN*

The polarizability tensors (Table SI.1) of *RM734* and *RM734-CN* were calculated at the M06HF-D3/cc-pVDZ level of DFT for a wavelength of 800 nm.

|  | λ (nm) | Iso ($C^2m^2J^{-1}$) | Aniso ($C^2m^2J^{-1}$) | Eigenvalues of the static polarizability tensor ($C^2m^2J^{-1}$) |
|---|---|---|---|---|
| RM734 | 800 | 44.7 | 45.7 | {75.0844, 43.3846, 19.6424} |
| RM734-CN | 800 | 46.4 | 50.8 | {77.7493, 41.9472, 19.5296} |

**Table SI.1:** Polarizabilities of *RM734* and *RM734-CN* as calculated using M06HF-D3/cc-pVDZ level of DFT.

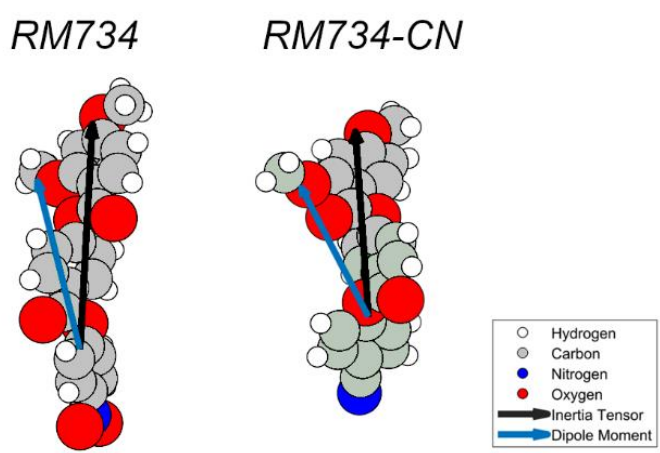

**Figure SI.5:** Directions of the inertia tensor and dipole moment vector for *RM734* and *RM734-CN* as calculated using M06HF-D3/aug-cc-pVTZ level of DFT**.**

# E. Conformational distributions calculated from MD simulations

We extracted the conformational distributions of several dihedrals from MD simulations of *RM734* and *RM734-CN* in both polar and apolar configurations at simulation temperatures of 400K as a means to complement torsional potentials calculated at the DFT(M06HF-D3/aug-cc-pVTZ) level which are discussed in the manuscript. Dihedral angles were calculated from the atomic coordinates over the full production MD trajectory and were binned into histograms to give the plots shown below. Differences in conformational populations between the two materials and two configurations are minor.

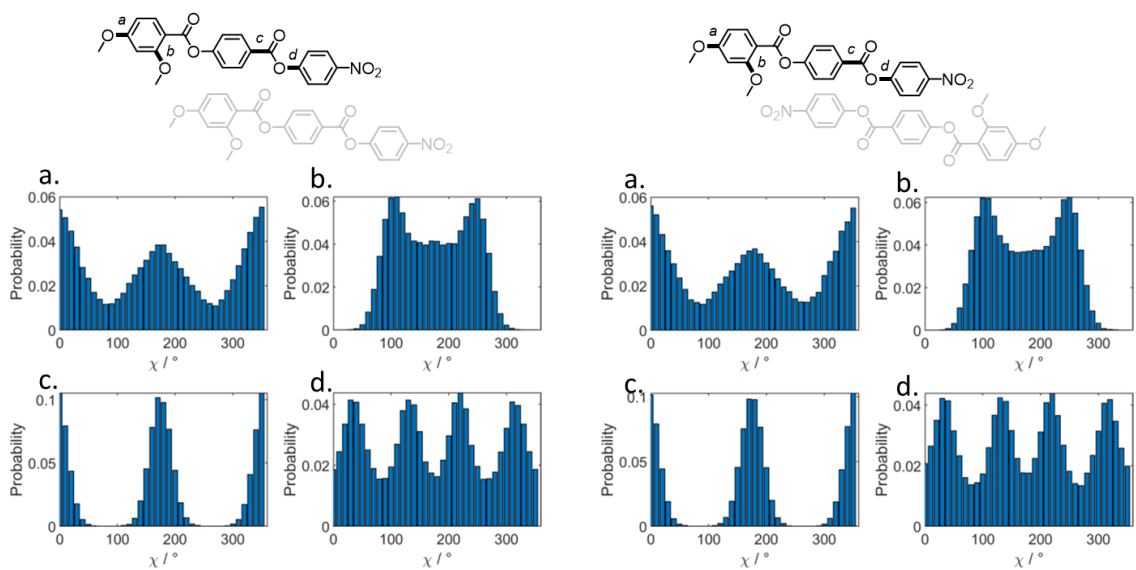

**Figure SI.6:** Conformational distributions from an MD simulation of *RM734* at 400K in the (left) polar nematic configuration and (right) apolar nematic configuration.

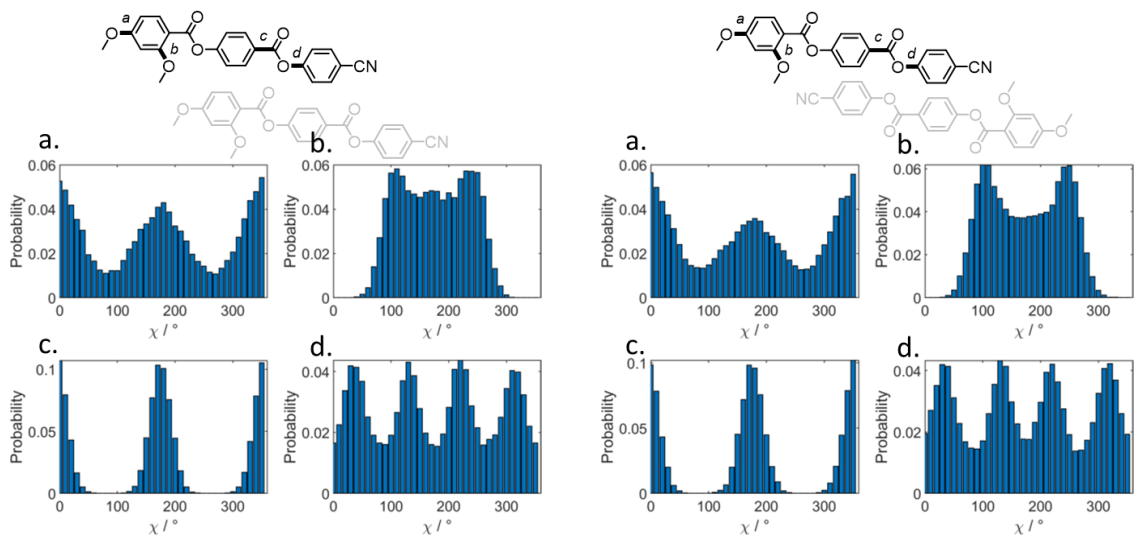

**Figure SI.7:** Conformational distributions from an MD simulation of *RM734-CN* at 400K in the (left) polar nematic configuration and (right) apolar nematic configuration.

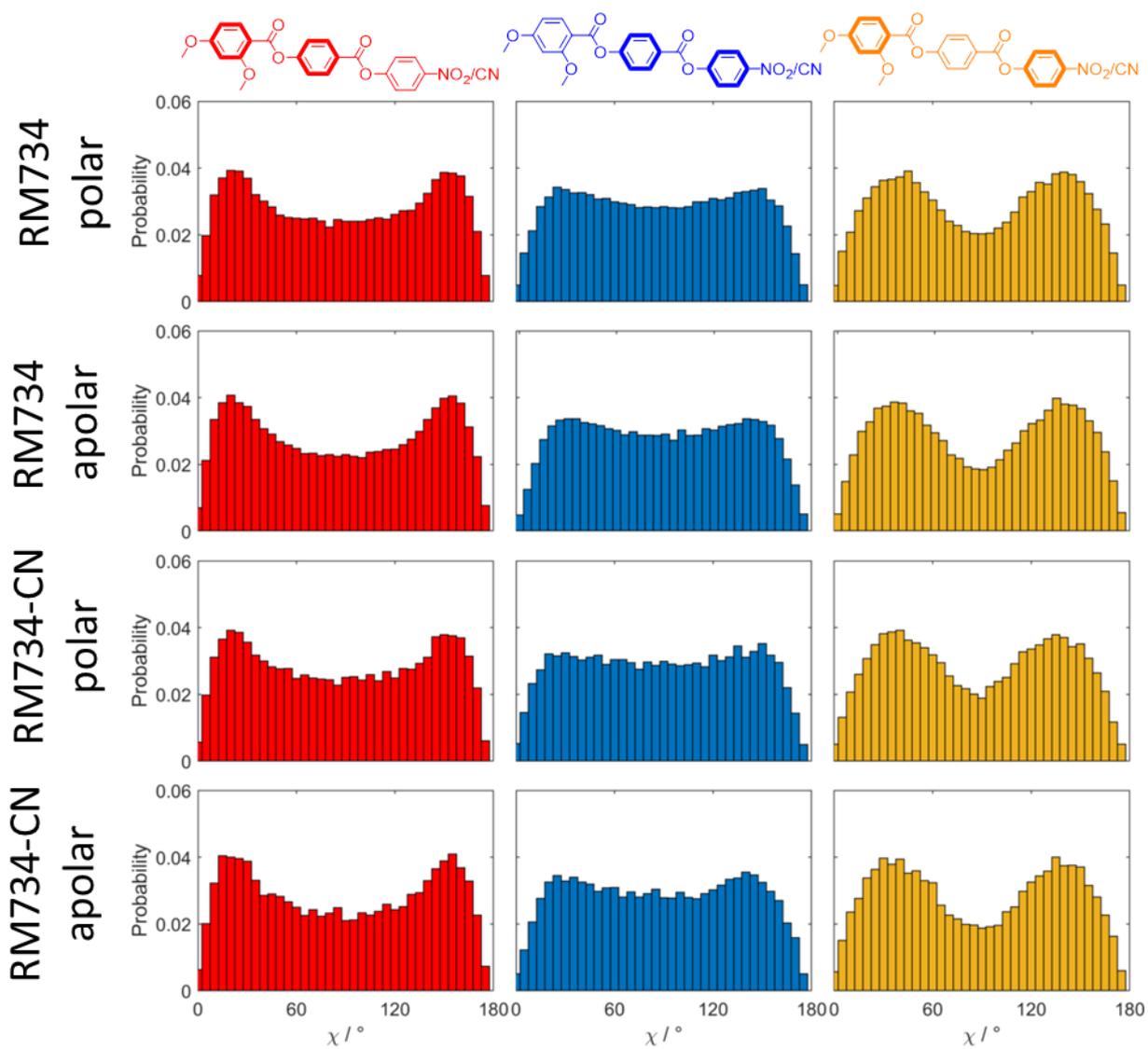

**Figure SI.8:** Angular distribution between the planes of the aromatic rings at 400K of *RM734* and *RM734-CN* in the polar and apolar configuration.

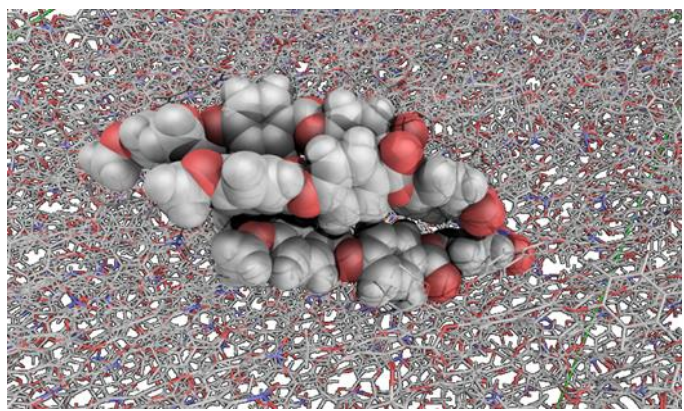

**Figure SI.9:** Visualization of staggered positions for *RM734* molecules in the polar configuration as represented from MD trajectories.

## F. Experimental and calculated scattered intensities for RM734

For each MD simulation, we calculated two-dimensional WAXS patterns as an average of trajectories in the time window 200 – 280 ns, as described in the experimental section of the manuscript. For the resulting 2D WAXS patterns we then calculated orientational order parameters as described elsewhere. Although this is an unconventional way to obtain orientational order parameters from MD simulations, it makes the same assumptions as used for experimental data and so provides directly comparable results. Reassuringly, values obtained from simulated WAXS patterns are not significantly different from the values obtained directly from MD simulations at the same temperature; for *RM734* we obtain *P2* values of 0.78 and 0.73 in the polar and apolar states, respectively, and for *RM734-CN* we obtain *P2* values of 0.75 and 0.74 in the polar and apolar states, respectively.

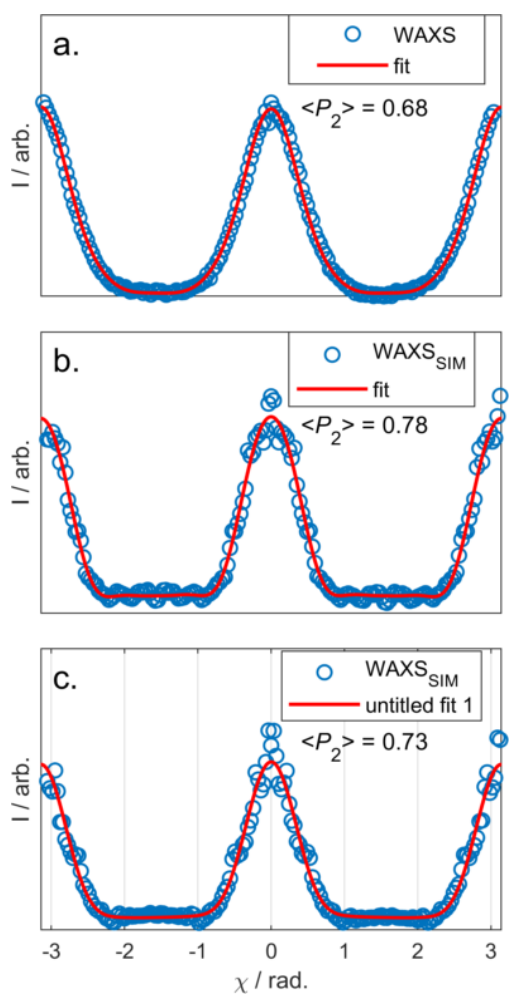

**Figure SI.10:** Scattered intensity as a function of χ for radially integrated wide-angle X-ray scattering patterns of *RM734*: (a) experimental data from reference 22 from the main manuscript; simulated data for the polar (b) and apolar (c) nematic configurations, obtained by azimuthal integration of simulated 2D WAXS patterns in the Q range 1 – 1.6 Å$^{-1}$. In all three examples data were fitted to $I(X) = \sum_{0}^{\infty} \frac{\pi}{2} f_{2n} \frac{(2n-1)!!}{2^n n!} cos^{2n}X$, and <*P2*> calculated as $\langle P_2 \rangle = \frac{1}{2}(3\langle cos^2\beta \rangle - 1)$, where $\langle cos^2\beta \rangle = \sum_{n=0}^{\infty} \frac{f_{2n}}{2n+3} \Big/ \sum_{n=0}^{\infty} \frac{f_{2n}}{2n+1}$

## G. Energy per molecule as calculated from MD

Energy per molecule is calculated from the total simulation energy (average over the whole production MD trajectory) and divided by the number of molecules. One should keep in mind that numbers are approximate and typically only relative changes are meaningful. Consecuently comparisons between materials should be avoided. Relative energy difference between the polar and the apolar configurations shows that for *RM734* polar configuration is more favourable. In the case of *RM734-CN*, the difference between both configurations is negligible.

**Table SI.2:** MD simulation total energies and energy per molecule at 400K of *RM734* and *RM734-CN*.

|  | Configuration | E Total (kJ/mol) | E molecule (kJ/mol) | $\Delta E$ molecule (kJ/mol) |
|---|---|---|---|---|
| *RM734* | POLAR | -53189.6 | -78.22 | 2.19 |
|  | APOLAR | -51700.4 | -76.03 |  |
| *RM734-CN* | POLAR | -371396 | -546.17 | 0.14 |
|  | APOLAR | -371304 | -546.03 |  |

## H. RM554

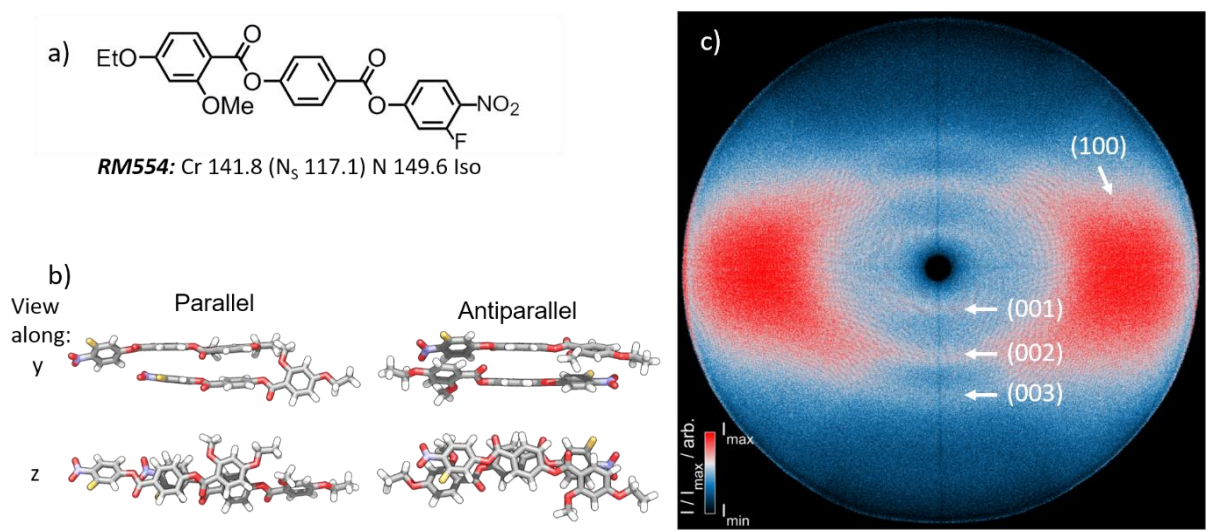

**Figure SI.11:** *RM554*. (a) Molecular structures and transition temperatures (°C), (b) Counterpoise corrected geometries of homogeneous dimer pair of *RM554* viewed along the y and z axes. All calculations were performed at the M06HF-D3/aug-cc-pVTZ level of DFT, as implemented in Gaussian G09 revision d01. (c) Magnetically aligned two dimensional X-ray scattering patterns in the $N_S$ phase at 87 °C. X-ray pattern shows the wide angle (100) and low angle (001) scaterring peaks characteristic of classical nematic materials and exhibits additional diffuse small angle reflections (002) and (003) similarly to *RM734*.

| Cpd. | $\omega_{initial}$, ° | $\omega_{final}$, ° | $\Delta E_{int}$, kcal mol$^{-1}$ | d(NO$_2$-COO)* |
|---|---|---|---|---|
| *RM734* | 0 | 8 | -19.96 | 3.55 |
|  | 180 | 165 | -34.32 | 12.89 |
| *RM734-CN* | 0 | 27 | -18.12 | 4.47 |
|  | 180 | 166 | -31.39 | 11.81 |
| *RM554* | 0 | 8 | -21.35 | 3.52 |
|  | 180 | 167 | -35.39 | 10.73 |

**Table SI.3:** Counterpoise corrected complexation angles and energies ($E_{int}$, Kcal mol$^{-1}$) for homogenous parallel and antiparallel dimers of *RM554* compared with those reported in the manuscript for *RM734* and *RM734-CN*. Distances (Å) between 'N' of -NO$_2$ or 'C' of -CN and 'C' of –COO- group are also given. All calculations were performed at the M06HF-D3/aug-cc-pVTZ level of DFT, as implemented in Gaussian G09 revision d01.

## I. Solid State of RM734

We revisited the solid-state crystal structure of *RM734*, which is available from the CCDC as deposition number 1851381. We find that, in agreement with DFT calculations for parallel pairs of *RM734* (and also *RM554*, above), there is a 'close contact' between the carbon atom of the ester (COO) group and oxygen atom of the nitro (NO$_2$) group, with a distance of 3.33 Å (Fig. SI-12). This distance is about 0.1 Å larger than the sum of the VDW radii of the atoms.

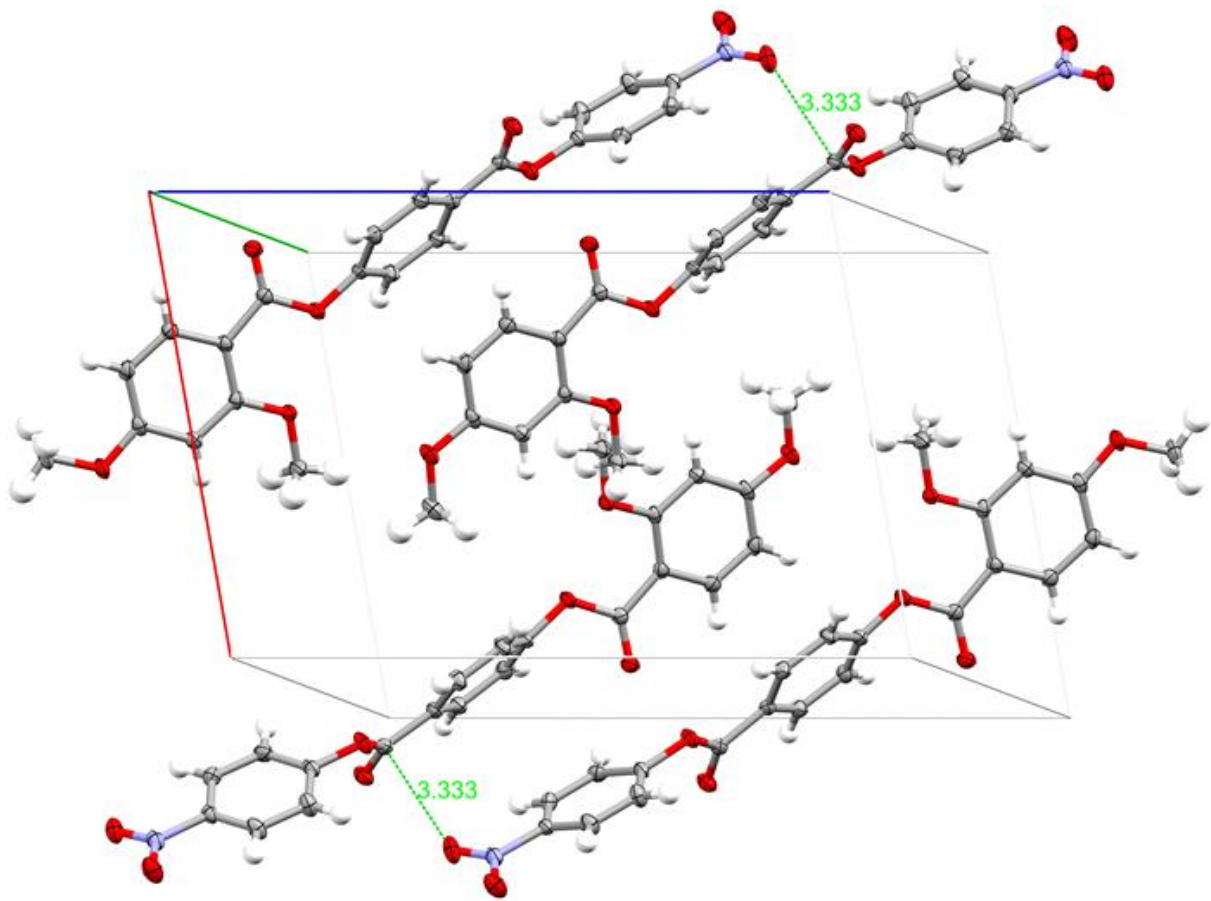

**Fig SI.12:** Structure of RM734 displayed as a thermal ellipsoid model (50% probability), obtained via X-ray diffraction. The unit cell (space group $P\bar{1}$) is indicated. Green lines correspond to nitro-ester close contacts, as described in the text. Viewed perpendicular to the AC plane, along the reciprocal B axis.